\documentclass[journal]{IEEEtran}
\IEEEoverridecommandlockouts
\makeatletter
\def\ps@headings{%
\def\@oddhead{\mbox{}\scriptsize\rightmark \hfil \thepage}%
\def\@evenhead{\scriptsize\thepage \hfil \leftmark\mbox{}}%
\def\@oddfoot{}%
\def\@evenfoot{}}
\makeatother
\usepackage{subfigure}
\usepackage{colortbl}
\usepackage{bm}

\usepackage{graphicx}  % Written by David Carlisle and Sebastian Rahtz
\usepackage{url}       % Written by Donald Arseneau

\usepackage{amsmath}   % Based on the American Mathematical Society
\usepackage{cite}

\usepackage{amsfonts,amssymb}

\usepackage{stfloats}

\usepackage{cases}
\usepackage{algorithm}
\usepackage{multirow}
\usepackage{algorithmic}
\usepackage{stfloats}%%%% �Ҽӵ�
\usepackage{epstopdf}%LRY ��
\usepackage{esint}
\usepackage{xcolor}
\usepackage{xpatch}
\makeatletter
\def\changeBibColor#1{%
	\in@{#1}{Non-coherent-RIS1, Non-coherent-RIS2, Non-coherent-RIS3}
	\ifin@\else\normalcolor\fi
}
\xpatchcmd\@bibitem
{\item}
{\changeBibColor{#1}\item}
{}{\fail}
\xpatchcmd\@lbibitem
{\item}
{\changeBibColor{#2}\item}
{}{\fail}

\newcommand{\Rmnum}[1]{\expandafter\@slowromancap\romannumeral #1@}

\makeatother

\newtheorem{remark}{{Remark}}

\newcommand{\ls}[1]
    {\dimen0=\fontdimen6\the\font
     \lineskip=#1\dimen0
     \advance\lineskip.5\fontdimen5\the\font
     \advance\lineskip-\dimen0
     \lineskiplimit=.9\lineskip
     \baselineskip=\lineskip
     \advance\baselineskip\dimen0
     \normallineskip\lineskip
     \normallineskiplimit\lineskiplimit
     \normalbaselineskip\baselineskip
     \ignorespaces
    }

\begin{document}

\title{Location-Aided Distributed Beamforming \\for Near-Field Communications\\ with Element-Wise RIS}
\vspace{10pt}

\author{
\IEEEauthorblockN{Xiao Zheng, \emph{Graduate Student Member}, \emph{IEEE}, Wenchi Cheng, \emph{Senior Member}, \emph{IEEE}, Jingqing Wang, \emph{Member}, \emph{IEEE}, Zhuohui Yao, \emph{Member}, \emph{IEEE}, and Jiangzhou Wang, \emph{Fellow}, \emph{IEEE}}

%\IEEEauthorblockA{$^{\dagger}$State Key Laboratory of Integrated Services Networks, Xidian University, Xi'an, China\\$^{\ddagger}$School of Engineering, University of Kent, Canterbury, United Kingdom\\
%E-mails: \{\emph{zheng\_xiao@stu.xidian.edu.cn, wccheng@xidian.edu.cn, j.z.wang@kent.ac.uk}\}}

%\vspace{-25pt}
\thanks{Part of this work was presented in IEEE Global Communications conference, 2024\cite{Xiaozheng_conf}.

This work was supported in part by the National Natural Science Foundation of China under Grant 62341132 and Grant QTZX24078, and also in part of the Fundamental Research Funds for the Central Universities (ZYTS25272). \emph{(Corresponding authors: Wenchi Cheng.)} 

Xiao Zheng, Wenchi Cheng, Jingqing Wang, and Zhuohui Yao are with Xidian University, Xi'an, 710071, China (e-mails: zheng\_xiao@stu.xidian.edu.cn; wccheng@xidian.edu.cn; jqwangxd@xidian.edu.cn).

Jiangzhou Wang is with the School of Information Science and Engineering, Southeast University, and Purple Mountain Laboratories, Nanjing 211119, China (e-mail: j.z.wang@kent.ac.uk).

}
}

\maketitle

\begin{abstract}
Active reconfigurable intelligent surface (RIS) emerges as an effective technique to resist the double-fading attenuation of passive RIS. By embedding with power harvesting function, it further evolves to zero-power active RIS, which can effectively enhance the flexibility of RIS deployment without external power demand. Nevertheless, existing works neglected the inherent difficulty of channel estimation (CE) for RIS-assisted systems, and the discrete phase shift constraint in practical deployment. In this paper we design a new element-wise RIS architecture and propose a distributed location-aided transmission scheme with low complexity to enhance the reflected gain for channel state information (CSI)-limited RIS-assisted near-field communications. Specifically, the new element-wise RIS provides dynamic element selection capability with low hardware resources. Based on Fresnel diffraction theory, we construct the mapping from locations in space-domain to phase distributions of waves in phase-domain and reveal the priority of elements for harvesting and reflecting. {Then, the distributed beamforming design with the phase of determine-then-align is proposed, where the estimation overhead reduction stems from exempted requirements of RIS-associated CE at base station (BS).} The asymptotic analysis indicates that the proposed scheme can achieve the optimal gain with a fixed proportion of reflective elements when RIS is large, followed by simulations to verify its superiority to other protocols.
\end{abstract}

\vspace{5pt}


\section{Introduction}
\IEEEPARstart{I}{n} light of great channel deterioration due to the high pathloss and weak penetrability of wave propagation for six-generation (6G) frequency bands, reconfigurable intelligent surface (RIS) with low-cost and energy-efficient passive elements can construct an additional reflected link to improve the transmission condition by reconfiguring the radio environment\cite{RIS-intro3,RIS-intro2,Harvest3}. However, it suffers from double-fading attenuation, leading to limited reflected gain\cite{Active_RIS1}. To combat this defect, the active RIS, endowing the capability of incident signal amplification by connecting elements with power amplifiers, can effectively improve the reflected gain without much power consumption\cite{Active_RIS2,Active_RIS3,RIS22}. {Regardless of the superiority of active RIS in performance improvement, achieving the desired gain in practical systems greatly limits its flexibility. Although it necessitates much less power budget than a relay, the power required by connected power amplifiers greatly exceeds that of a passive RIS, especially for large-size RISs with high phase shift resolution\cite{Active_RIS5,Active_RIS6}. To ensure stable and long-term reflected gain, it necessitates external power supply to enable the signal amplification function considering the non-negligible power budget of active RIS, typically by cables, thus greatly shrinking the deployable area\cite{AutonomousRIS}.}
 
{To address this problem, new RIS architecture embedded with both energy harvesting (EH) and signal reflection functions emerges to provide sustainable communication enhancement capability. For example, the authors of \cite{Harvest2} proposed a novel concept of multi-layer refracting RIS assisted receiver to achieve concurrent transmission of the information and energy, and overcomed the severe fading effect induced by extreme long-distance links by fully exploiting RIS's degrees-of-freedom.} The authors in \cite{Selfris4} designed a power splitting scheme (PS), where a fraction of energy from the incident signal is harvested while the remaining energy is reflected. The authors of \cite{Selfris3} developed the joint beamforming scheme with the time-switching (TS) strategy by alternating optimization algorithm. The authors in \cite{Selfris6} analyzed the performance of self-sustainable RIS assisted system under the perfect and imperfect channel state information (CSI) cases, respectively. In \cite{Harvest3}, the authors proposed a penalty-function based dual decomposition scheme to achieve the balance among transmit power budget, harvested power, and signal reception constraints by absorptive RIS with PS strategy. The authors in \cite{Selfris1} explored the performances of three prominent protocols with respect to zero-energy RIS including PS, TS, and element-splitting (ES) and derived the corresponding theoretical outage probabilities, respectively. Utilizing harvest-then-reflect design, the essential power consumption budget for RIS's controller and amplifying circuits can be completely self-sufficient by the harvested power, which exhibits superior performance to passive RIS and can greatly enhance deployment flexibility.

Nevertheless, the considered scenario is ideal in these works, which neglected the inherent difficulty of channel estimation (CE) for RIS-assisted systems, and the discrete phase shift constraint. { The priori knowledge of CSI mainly impacts the detailed system design from the perspective of both transmitter and receiver. At receiver, the CSI is important to determine the performance of signal detection. To release the limitation of CSI acquisition, the authors in \cite{Non-coherent-RIS1,Non-coherent-RIS2,Non-coherent-RIS3} make great advances in non-coherent detections for RIS-assisted system, where the receiver can retrieve information bits by performing non-coherent correlation demodulation without requiring CSI. At transmitter, however, the priori information about channel parameters of RIS-associated links is imperative to determine the configuration profile of RIS, requiring complicated CE design.} {Especially, considering the enlarged RIS's aperture and ascending operation frequency in 6G, the near-field propagation with spherical wave condition typically holds for wireless communications. For example, the Rayleigh distance is 240 m when the RIS area in square shape is 1 $\rm m^2$ when operating in 100 GHz, which typically covering the 5G NR cell. Furthermore, the near-field range in RIS-assisted MIMO systems is further expanded by jointly considering the BS's aperture and UE's aperture as well as the distances of BS-RIS and RIS-UE links \cite{Near-Field-MIMO}. Once one of the distances is less than the near-field region, the overall propagation should be considered as spherical wave to finely capture the phase distribution. Hence, investigating the near-field propagation in RIS-assisted communications is of vital significance.} The non-linear spherical array response embedded with both angle and distance information aggravates the difficulty of CE, further increasing large complexity and overhead, which is adverse to exploiting the favorable reflected gain of RIS\cite{near_field_Challenge}. Additionally, in light of hardware constraint for RIS's fabrication, such as positive-intrinsic-negative (PIN) diodes, the achievable phase shift is generally discrete, leading to the presence of phase control deviations in practical systems\cite{discreteRIS1}. Such deviations exhibit an element-wise distribution with respect to channel condition. However, most of existing works overlooked this effect and considered the PS and TS strategies to balance the power consumption and EH of RIS\cite{Selfris2,Selfris3,Selfris4,Selfris6,Harvest1}. Although the authors in \cite{Selfris1} explored the performances of ES strategy by dividing part of elements for EH and others for reflection, the dynamic element selection scheme with respect to channel condition has not been investigated.

Fortunately, in RIS-assisted near-field communications, high line-of-sight (LoS) probability and ultra-precision positioning can be supported\cite{Spherical_Localization}. {Capturing the promoted sensing accuracy, e.g., fine-grained location information, has the potential to release the complex interactive procedure of CE by utilizing the priori information with much less complexity, where sensing-assisted beamforming can significantly promote the effectiveness of beamforming based on sensing results\cite{Location_aided1,Location_aided2}. As opposed to CSI-enabled beamforming schemes which necessitate independent pilot signals to estimate RIS-associated CSI, exploiting the location for beamforming design can potentially decrease pilot overhead and computational complexity.} In specific, the pilot overhead can be reduced due to sharing capability of the estimation target among RISs regarding one user equipment (UE)\cite{Xiaozheng_mag}. Hence, exploiting locations as the priori information for beamforming design can greatly reduce the pilot overhead from proportional to the number of RIS to proportional to the number of users. Meanwhile, RIS-assisted systems under CSI-driven schemes often operate within a centralized framework, where precoding and phase shift design are typically controlled by base station (BS) versus dedicated control channels and explicit control signaling. Such framework heavily depends on CE of RIS-associated channels, resulting in enormous complexity. Therefore, the decentralized frameworks bypassing dedicated and explicit control is a favorable alternative for future RIS-assisted systems\cite{Autonomous_RIS,Autonomous_RIS2,AutonomousRIS}.

{
Inspired by the above observations, in this paper we design a new element-wise RIS architecture based on element selection and propose a distributed location-aided transmission scheme in the presence of discrete phase shift constraint for RIS-assisted near-field communications. By harnessing the location information, such design can support the RIS controller to dynamically determine the mode of elements according to their priorities in phase distribution, which can alleviate the hard requirement of RIS-associated CSI and further improve the overall performance by capturing the element-wise gain. {The main contributions of this work are summarized as follows:}

\begin{itemize}
    \item Considering the discrete phase resolution in practical deployment, we propose a new element-wise RIS architecture, which endows RIS with the self-sufficiency and location acquisition capabilities. By theoretically analyzing the phase deviations induced by hardware constraints, we show that the phase deviations of RIS elements greatly affect their effective gains, exhibiting different priorities for harvesting and reflecting. We exemplify with 1-bit and 2-bit phase resolution cases to explicate its inherent mechanism, thus extending to the element selection criterion based on the phase distribution of wave and universal phase resolution of RIS. 
    \item We exploit location information to characterize the propagation phase distribution by Fresnel diffraction theory. By solving the formulated parametric ellipsoid equations, the mapping from locations in space-domain to phase distributions of waves in phase-domain is constructed. Then, we detail the proposed frame and distributed transmission scheme. Therein, the estimation and beamforming stages are decentralized to reduce the estimation complexity and overhead in RIS-assisted communications. A suboptimal but lightweight solution with slight performance loss to the complex optimization problem is derived by the determine-then-align design.
    \item We provide theoretical analysis to evaluate the achievable gain by the proposed RIS design compared with passive RIS. The asymptotic result indicates that the optimal proportion of reflective elements approaches a fixed value when RIS is large to provide the maximum gain. Also, the complexity analysis reveals that the complexity of the distributed beamforming scheme linearly increases as the number of RIS elements increases. Finally, simulation results show that the performance of the proposed element-wise design surpasses that of the traditional TS and EW strategies, while exhibiting competitive performance to that of the PS strategy with low phase resolution, verifying its superiority in practical deployment. 
\end{itemize}
}

The remaining of this paper is organized as follows: Section II introduces the system model of the RIS-assisted near-field communications consisting of the proposed element-wise RIS structure and the channel model, followed by optimization problem formulation. Section III constructs the mapping from locations to phase distributions by Fresnel zone model and elucidates the behavior of phase deviations. {In Section IV, the distributed transmission design including the frame structure, estimation, and beamforming design is presented, followed by the complexity and asymptotic gain of the proposed scheme.} Simulations are presented in Section V and finally Section VI concludes the paper. 
\begin{figure*}[t]
    \centerline{\includegraphics[width=16cm]{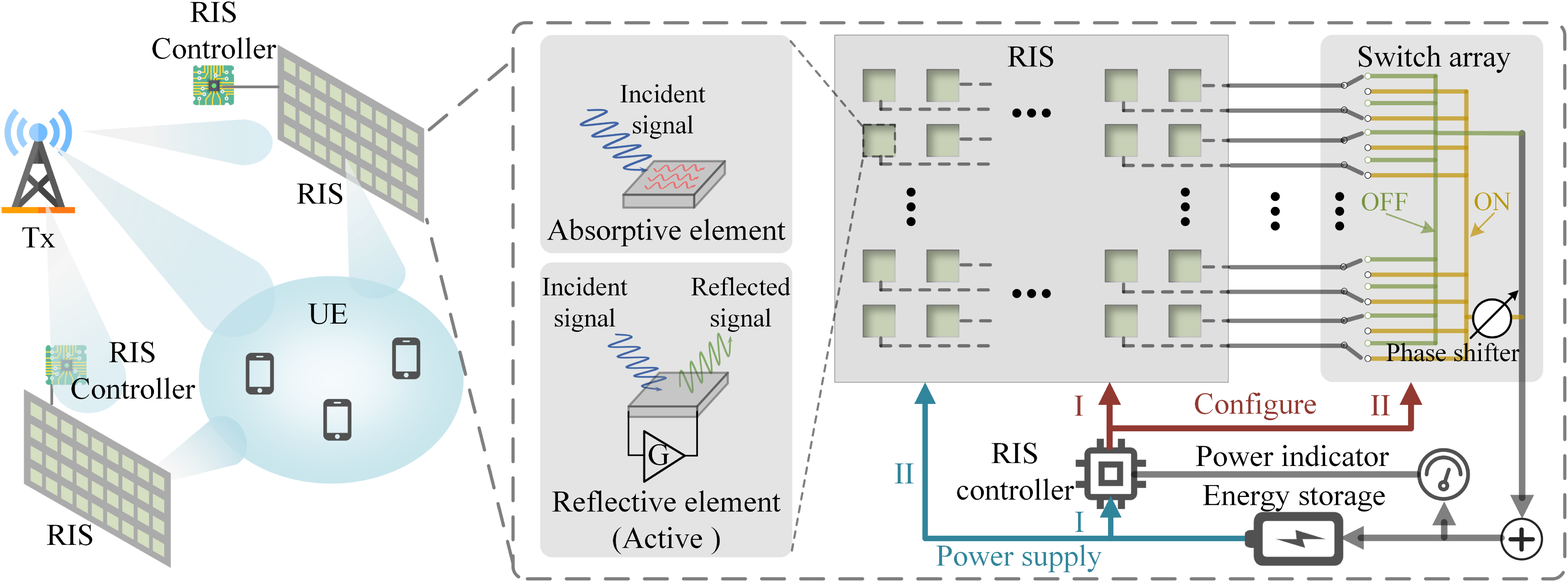}}
        \caption{The system model for self-sustainable active RIS assisted near-field communications.}
        \label{fig1}
\end{figure*}

{\emph{Notations:} $\left[\cdot\right]^{\rm H}$ denotes the conjugate-transpose operations; $\left\lVert\cdot \right\rVert $ denotes the Euclidean norm of the argument; $\left\lVert\cdot \right\rVert _{\rm F}$ denotes the Frobenius norm of the argument; diag$(\cdot)$ denotes the diagonalization operation; $\varepsilon\left(\cdot\right)$ denotes the step function; $\text{mod}\left(a,b\right)$ denotes the modulo operation; $\odot$ denotes the Hadamard product; $\angle$ denotes the phase of vector; $\text{exp}\left(\cdot\right) $ denotes the exponential function.}

\section{System Model}
{Figure~\ref{fig1} depicts the system model of location-aided near-field communications surrounded by multiple large RISs with both EH and active reflective functions.} {Therein, a BS equipped with $N_T$ antennas is assisted by $K$ RISs equipped with $N_R$ elements to serve user equipments (UEs) equipped with single antenna.} Generally, the RISs are deployed at fixed positions to provide virtual LoS paths. {Therefore, it is assumed that the locations of BS antennas can be perfectly known at the RIS controllers.} {Also, a single antenna is unable to generate directional beam. Hence, each UE's antenna is assumed to be isotropic (omnidirectional).} The location of the $n_t$-th antenna of the BS and the location of the $n_r$-th element of the $k$-th RIS in the 3D Cartesian coordinate system are denoted by ${\boldsymbol{\rm  t}}_{n_t}=\left[t_{n_t,x},t_{n_t,y},t_{n_t,z}\right] $ and $\boldsymbol{\rm  r}^k_{n_r}=\left[r^k_{n_r,x},r^k_{n_r,y},r^k_{n_r,z}\right] $, respectively, with $n_t=1,\cdots, N_T$, $k=1,\cdots, K$, and $n_r=1,\cdots, N_R$, respectively. During the transmission, the time division protocol is adopted where each time slot resource is allocated to one UE. For generality, due to the mobility of UE and the inevitable environment noise, localization accuracy of UE is always limited, resulting in location uncertainty. {Hence, the location of the UE, denoted by $\boldsymbol{\rm u}=\left[u_{x},u_{y},u_{z}\right] $, is assumed to be impacted by the general 3D Gaussian localization noise\cite{Gaussian_noise}, denoted by $\boldsymbol{\rm  n}_l=\left[n_{l,x},n_{l,y},n_{l,z}\right] \thicksim \mathcal{N}(\boldsymbol{0}_3,\sigma_l^2 \boldsymbol{\rm I}_3)$ with $\boldsymbol{\rm I}\in\mathbb{C}^{3\times 3}$ being the unit matrix. }

\subsection{Element-Wise RIS Model}
We commence by encapsulating basic functions of the element-wise RIS model. {As depicted in the right part of Fig.~\ref{fig1}, each RIS comprises massive two-state elements, a RIS controller, a switch array with phase shifter, a power indicator, and an energy storage module. Such items enable the RIS with three main functions: \textbf{self-sufficiency}, \textbf{element-wise control}, and \textbf{location acquisition}. First, each element can be configured to absorptive or active reflective modes, {where absorptive elements can fully harvest the energy of incident signals and convert them into electrical power}, followed by charging the energy storage module, e.g., rechargeable battery. {The reflective element is active and embedded with power amplifiers to amplify then reflect the incident signals with controllable phase shift.} Such design gives RIS the capability of self-sufficiency for the power consumption of these active elements.} {Second, the element-wise selection capability is realized by the switch array module, which contains $2N_R$ wires and $N_R$ dual selector switches to gather incident signals with the same interference condition\footnote{In the proposed design, the number of ports of selector switches is set to 2 to distinguish two streams. Then, by applying one phase shifter, the two streams denoted by ON and OFF states composed of signals with the same constructive and destructive interferences can be coherently aligned to maximize the energy of the composite stream. Compared with fully-digital design, the proposed design can greatly reduce the number of phase shifter from $N_R$ to 1 at the expense of slight loss of gain.} (constructive and destructive interferences).} {Third, the composite stream is followed by a power indicator. By exploiting the spherical wavefront of the near-field propagation, UE's location can be extracted by proper absorptive coefficients design to capture the sensing capability at the RIS side.} Then, the harvested energy is fed into the energy storage module, which is used to supply power for the entire RIS operation mainly including two parts: I) circuit consumption of the RIS controller; II) the power supply feeding to active reflective elements to enhance the reflected performance. By dynamically configuring the proportion of different types of elements, the new RIS architecture can be completely self-sufficient without the demand for external power feed. The RIS controller receives control information from BS, and correspondingly configures two modules: I) control the switch array to enable the element-wise design by selecting the state of RIS element; II) determine the reflective and absorptive coefficients of RIS elements according to detailed demands.
 
The reflective coefficient matrix of the $k$-th RIS is given by $\boldsymbol{\Phi}^{\rm r}_k={\rm Diag}(\varrho _{k,1}^{\rm r},\cdots,\varrho _{k,N_R}^{\rm r})\in \mathbb{C}^{N_R\times N_R}$, where the reflective coefficient of the $n_r$-th active element is represented by $\varrho _{n_r}^{\rm r}=\rho_{n_r}^re^{j\theta_{n_r}}$ with $n_r=1,\cdots, N_R$. Therein, $\rho_{n_r}^r$ and $\theta_{n_r}$ denote the amplitude and phase shift of the $n_r$-th reflective coefficient, respectively, with $\rho_{n_r}^r \in \left\{0, \rho \right\} $ and $\theta_{n_r}\in \left[0, 2 \pi\right) $, respectively. Here, without loss of generality, it is assumed that $\rho$ denotes the fixed magnification of amplifying circuits determined by amplifiers and all reflective elements share the same magnification\footnote{{Generally, the power of incident wave on individual RIS element is very limited. Hence, in this paper we only consider the linear amplifying region where the power of reflected wave is determined by $\rho$. Also, to reduce hardware overhead, the cascade solution is typically adopted where amount of elements share one amplifier with the same amplifying factor. Hence, in this paper, we assume that the amplifying factors of all reflective elements are the same for simplicity.}}, typically larger than 1. Notice that the reflective coefficient of absorptive elements is set to 0 as absorptive elements can completely absorb incident signals without reflection. Meanwhile, we define $\boldsymbol{\Phi}^{\rm a}_k={\rm Diag}(\varrho _{k,1}^{\rm a},\cdots,\varrho _{k,N_R}^{\rm a})\in \mathbb{C}^{N_R\times N_R}$ as the absorptive coefficient matrix of the $k$-th RIS, where $\varrho _{k,n_r}^{\rm a}\in \left\{\varrho ^{\rm a}, 0\right\}$ represents the absorptive efficiency of the $n_r$-th RIS element with the same maximum efficiency for $n_r=1,\cdots,N_R$ and $k=1,\cdots,K$. {In addition, in this paper we mainly focus on the phase design of the joint reflective and absorptive element so that the coupling effect of RIS is ignored\footnote{In practical hardware implementations, the phase and amplitude responses of the RIS are typically coupled, thus inevitably leading to a performance loss. In this paper, we mainly focus on the joint design of reflective and absorptive elements in the phase domain. The electromagnetic model of the proposed element-wise RIS architecture will be investigated in our future work by considering the coupling effect to extend its generality.}.}

More importantly, due to the practical hardware limitation and fabrication cost, the phase shift values of RIS are typically discrete. {Higher phase precision requires higher power consumption and more complicated circuits}\cite{discreteRIS1}. In this paper we consider a general model for RIS with discrete phase shifts as follows:
\begin{equation}
    \theta_{n_r}\in \mathcal{S} \overset{\triangle }{=}\left\{0, \frac{2\pi}{2^D},\cdots,\frac{2\pi\left(2^D-1\right)}{2^D}\right\},
\end{equation}
where $\mathcal{S}$ denotes the set of phase shift value with $\theta_{n_r}$ being equally valued from $\left[0, 2\pi\right) $ and $D$ denotes the discrete phase resolution of the RIS elements. 

\subsection{Channel Model}
Due to the expansion of the near-field region in RIS-assisted communications in high frequency bands, we use the typical near-field channel model, where the channel gain is highly related to the relative locations among BS, RISs, and UE. As shown in Fig.~\ref{fig1}, the communication channels between the BS and UEs consist of the direct channel and the RIS-assisted cascaded channel\footnote{As the power gain provided by RIS is limited, in this paper we only consider the primary reflected channel and omit the multi-reflection between RISs.}. Therein, the direct channel is denoted by $\boldsymbol{\rm f}\in \mathbb{C}^{N_T\times 1}$ with its $n_t$-th element being
\begin{equation}
    \begin{aligned}
        \boldsymbol{\rm f}\left({n_t}\right) =\frac{\sqrt{G_T\left(\theta^{TU}_{T},\phi^{TU}_{T}\right) G_U\left(\theta^{TU}_{U},\phi^{TU}_{U}\right) }\lambda}{4\pi\left\lVert \boldsymbol{\rm  t}_{n_t}-\boldsymbol{\rm  u}\right\rVert} \\\times{\rm exp}\left({-j\frac{2\pi}{\lambda}\left\lVert \boldsymbol{\rm  t}_{n_t}-\boldsymbol{\rm  u}\right\rVert }\right)+f^{\text{NLoS}}_{n_t}
    \end{aligned}
\end{equation}
according to the near-field path loss model \cite{NEAR_channel}, where the first item represents the channel gain of LoS link and the second item $f^{\text{NLoS}}_{n_t}$ denotes the unified channel gain including all multi-path non-line-of-sight (NLoS) components. Additionally, $G_T\left(\theta,\phi\right)$ and $G_U\left(\theta,\phi\right)$ represents the transmit antenna gain and receive antenna gain, respectively, $\theta^{TU}_{T/U}$ and $\phi^{TU}_{T/U}$ are the elevation and azimuth angles of the BS-UE links relative to the facing direction of BS/UE antenna, respectively, and $\lambda$ is the radio wavelength. Generally, $G_T\left(\theta,\phi\right)$ and $G_U\left(\theta,\phi\right)$ can be typically set as ${\rm cos}^q\left(\theta\right)$ function to approximate the radiation pattern of transmit and receive antennas\cite{RIS_Path_Loss}, which can be expressed as follows:
\begin{equation}
    \begin{aligned}
        G_{T/U}\left(\theta,\phi\right) =\frac{4\pi {\rm cos}^q\left(\theta\right) }{\int_0^{2\pi}\int_0^{\frac{\pi}{2}}{\rm cos}^q\left(\theta\right)d\theta d\phi}.
    \end{aligned}
\end{equation}
 
The channel between BS and the $k$-th RIS is denoted by $\boldsymbol{\rm h}_k\in \mathbb{C}^{N_T\times N_R}$, and the channel between the $k$-th RIS and UE is denoted by $\boldsymbol{\rm g}_{k}\in \mathbb{C}^{N_R\times 1}$. Accordingly, the $\left(n_t,n_r\right) $-th item of $\boldsymbol{\rm h}_k$ and the $n_r$-th item of $\boldsymbol{\rm g}_{k}$ can be expressed by 
\begin{equation}
    \begin{aligned}
        \boldsymbol{\rm h}_k\left({n_t,n_r}\right) =\frac{\sqrt{G_T\left(\theta^{TR_k}_{T},\phi^{TR_k}_{T}\right)G_R\left(\theta^{TR_k}_{R_k},\phi^{TR_k}_{R_k}\right)}\lambda}{4\pi\left\lVert \boldsymbol{\rm  t}_{n_t}-\boldsymbol{\rm  r}^k_{n_r}\right\rVert} \\\times {\rm exp}\left({-j\frac{2\pi}{\lambda}\left\lVert \boldsymbol{\rm  t}_{n_t}-\boldsymbol{\rm  r}^k_{n_r}\right\rVert }\right) 
    \end{aligned}
\end{equation}
and 
\begin{equation}
    \begin{aligned}
        \boldsymbol{\rm g}_{k}\left({n_r}\right) =\frac{\sqrt{G_U\left(\theta^{UR_k}_{U},\phi^{UR_k}_{U}\right)G_R\left(\theta^{UR_k}_{R_k},\phi^{UR_k}_{R_k}\right)}\lambda}{4\pi\left\lVert \boldsymbol{\rm  r}^k_{n_r}-\boldsymbol{\rm  u}\right\rVert} \\\times {\rm exp}\left({-j\frac{2\pi}{\lambda}\left\lVert \boldsymbol{\rm  r}^k_{n_r}-\boldsymbol{\rm  u}\right\rVert }\right) ,
    \end{aligned}
\end{equation}
respectively, which denotes the channel gain between the $n_t$-th antenna of BS and the $n_r$-th element of the $k$-th RIS, and the channel gain between the $n_r$-th element of the $k$-th RIS and the UE, respectively. Therein, $G_R\left(\theta,\phi\right)$ is the radiation pattern of RIS element\cite{Xiao_zheng2}, $\theta^{TR_k}_{T/R_k}$ and $\phi^{TR_k}_{T/R_k}$ are the elevation and azimuth angles of the link between the BS and the $k$-th RIS relative to the radiation direction of BS/RIS, respectively, and $\theta^{UR_k}_{U/R_k}$ and $\phi^{UR_k}_{U/R_k}$ are the elevation and azimuth angles of the link between the UE and the $k$-th RIS relative to the radiation direction of UE/RIS, respectively. As the channel gain corresponding to multi-reflection is very limited, we omit the NLoS components for RIS-associated channel and only consider its LoS link.

\subsection{Problem Formulation}
Hence, the received signal at the UE, denoted by $y_u$, can be written as 
\begin{equation}
    \begin{aligned}
        y_u=\sum_{k=1}^{K}\left(\boldsymbol{\rm f}^{\rm H}+\boldsymbol{\rm g}_{k}^{\rm H} \boldsymbol{\Phi}^{\rm r}_k\boldsymbol{\rm h}_k^{\rm H}  \right) \boldsymbol{\rm w}x+\sum_{k=1}^{K}\boldsymbol{\rm g}_{k}^{\rm H} \boldsymbol{\Phi}^{\rm r}_k\boldsymbol{\rm n}_k +n_u
        ,
    \end{aligned}
\end{equation}
where $\boldsymbol{\rm n}_k\thicksim \mathcal{C} \mathcal{N}(\boldsymbol{0}_{N_R},\sigma_r^2 \boldsymbol{\rm I}_{N_R})$ and $n_u\thicksim \mathcal{C} \mathcal{N}(0,\sigma_u^2 )$ denote the noise item at the $k$-th RIS and UE, respectively, $\boldsymbol{\rm w}\in \mathbb{C}^{N_T\times 1}$ is the precoding vector at the BS satisfying $\boldsymbol{\rm w}^{\rm H}\boldsymbol{\rm w}= 1$, and $x$ is the transmitted signal. Hence, the signal-to-noise ratio (SNR) at the UE is 
\begin{equation}
    \gamma=\frac{\left\lVert \sum_{k=1}^{K}\left(\boldsymbol{\rm f}^{\rm H}+\boldsymbol{\rm g}_{k}^{\rm H} \boldsymbol{\Phi}^{\rm r}_k\boldsymbol{\rm h}_k^{\rm H}  \right)\right\rVert^2 }{\left\lVert \sum_{k=1}^{K}\boldsymbol{\rm g}_{k}^{\rm H} \boldsymbol{\Phi}^{\rm r}_k\boldsymbol{\rm n}_k +n_u\right\rVert^2 }.
\end{equation}
Correspondingly, the harvested signal at the $k$-th RIS, denoted by $y_k$, can be expressed as follows: 
\begin{equation}
    \begin{aligned}
        y_k=\boldsymbol{1}^{1\times N_R}\boldsymbol{\Lambda}_k\odot \boldsymbol{\Phi}^{\rm a}_k \boldsymbol{\rm h}_k^{\rm H} \boldsymbol{\rm x}+n_k
        ,
    \end{aligned}
\end{equation}
where $\boldsymbol{1}^{1\times N_R}$ represents the vector consisting of 1, $\boldsymbol{\Lambda}_k=\text{diag}\left(\varsigma_{k,1},\cdots,\varsigma_{k,N_R} \right) \in \mathbb{C}^{N_R\times N_R}$ denotes the configuration matrix of the switch array of the $k$-th RIS in Fig.~\ref{fig1}, and $n_k$ is the noise item at the $k$-th RIS. {Therein, $\boldsymbol{\Lambda}_k$ is to align the two streams of absorptive elements with low hardware overhead. Assisted by one phase shifter, the two streams composed of signals with the same constructive and destructive interferences can be coherently aligned to enhance the harvested energy.} Then, the stored power harvested by the $k$-th RIS has the following expression
\begin{equation}
    \begin{aligned}
        P_k= \eta_1 \left\lVert \boldsymbol{1}^{1\times N_R} \boldsymbol{\Lambda}_k \odot \boldsymbol{\Phi}^{\rm a}_k \boldsymbol{\rm h}_k^{\rm H} \boldsymbol{\rm x}+n_k\right\rVert ^2
        ,
    \end{aligned}
\end{equation}
where $\eta_1$ denotes the conversion efficiency of the charging process. The available power for supplying the amplifying circuits at the $k$-th active RIS, denoted by $P_{a,k}$, can be expressed as follows:
\begin{equation}
    P_{a,k}=\eta_2 P_k -P_c -P_{DC}-f(D)N_R, \label{power_con}
\end{equation}
where $\eta_2$ denotes the conversion efficiency of the discharging process, $P_c$ denotes the power consumption at the RIS controller, $P_{DC}$ denotes the total direct current biasing power consumption of active circuits. {Additionally, $f(D)$ denotes the power consumption with respect to the phase resolution $D$. The power dissipated at each unit depends on the fabrication, i.e., varactor diodes or PIN diodes. Generally, $f(D)$ is a fixed value and exhibits linear relation with respect to $D$ for varactor diodes and PIN diodes based elements, respectively\cite{RIS_energy_model}.} Hence, to maximize the achievable rate at the UE, the SNR of the received signal at UE should be maximized, where the optimization problem can be formulated as follows:
{\begin{subequations}\label{Problem1}
	\begin{alignat}{2}
		\textbf{\textit{P}1:}
		&~~~~~~~~~~~~~~~~~~~~~~~\mathop{\max}\limits_{\boldsymbol{\Phi}^{\rm a}_k, \boldsymbol{\Phi}^{\rm r}_k,\boldsymbol{\Lambda}_k}  \gamma \notag \\
		{\rm{s.t.}}:&1).\ \left\lVert \boldsymbol{\Phi}^{\rm r}_k\right\rVert_{\rm F}\leq P_{a,k},~\forall k=1, \cdots, K  ; \\
		&2).\ \left\lVert \varsigma_{k,n_r}\right\rVert=1,~\forall  k=1,\cdots,K;n_r=1,\cdots,N_R; \\
        &3).\ \varrho _{k,n_r}^{\rm a}\in \left\{\varrho ^{\rm a}, 0\right\},~\forall  k=1,\cdots,K;n_r=1,\cdots,N_R;\\
		&4).\  P_{a,k}>0,~\forall  k=1,\cdots,K.
	\end{alignat}
\end{subequations}}
{Therein, (11a) is the available power constraint; (11b) and (11c) are constrained by the configuration matrix of the switch array and amplifying circuits, respectively; (11d) is to make RIS completely self-sustainable.} Due to the constraints (11b) and (11c), the formulated problem \textbf{\emph{P}1} is non-convex, resulting in significant computational complexity when solved versus iterative optimization methods. To address this problem, we elaborate on the inherent behavior of waves corresponding to the proposed RIS model by Fresnel diffraction theory in the next section,{ which can be used to pursue an interpretable solution for the optimization problem.}

\section{Fresnel Zone Model}
To characterize the phase distributions of the reflected waves with crucial role for beamforming design, in this section we depict the mechanism of the Fresnel zone model and then extend the traditional Fresnel zone model to fractional Fresnel zone to make it suitable for the general $D$-bit phase shift case. More importantly, by theoretically analyzing the phase deviations induced by hardware constraints, we show that the phase deviations of RIS elements greatly affect their effective gains, exhibiting different priorities for harvesting and reflecting. To capture the potential gain by the above priorities, we formulate the Fresnel ellipsoid equations and derive the parametric curves in the presence of UE's localization uncertainty to construct the mapping from space domain to phase domain. 

%For an assembled RIS, the near field region (e.g. Fraunhofer distance), denoted by $F_d$, will increase as the radio frequency increases according to its definition that $F_d=2D^2/\lambda$ with $D$ and $\lambda$ being the largest length of RIS and wavelength, respectively. To derive desired reflected power at users, the area of RIS cannot decrease as the radio frequency wavelength decreases, since the total reflected power is proportional to the area of RIS\cite{RIS-size}. Thus, the near field region in mmWave bands will achieve a hundred of meters when the RIS area in square shape is 1 $\rm m^2$. Thus, in typical RIS-assisted mmWave communications, the transmitter and receiver are located in the Fresnel zone of RIS, where diffraction theory can be leveraged to model the reflection of wave impinging on RIS. 

\subsection{Parametric Fresnel Ellipsoid}
\begin{figure}[t]
    \centerline{\includegraphics[width=8.2cm]{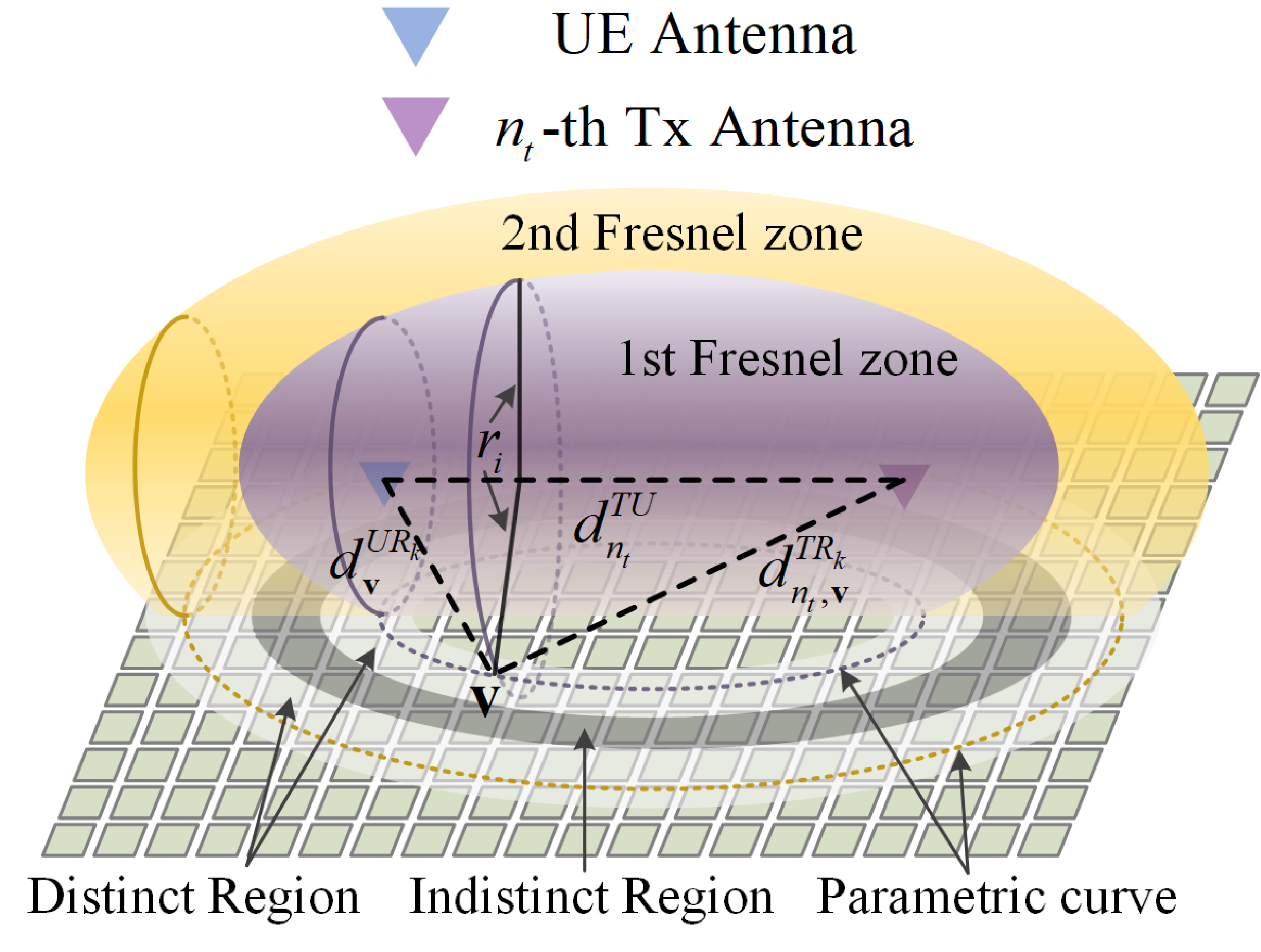}}
        \caption{The diagram of the first and second Fresnel Zones.}
        \label{fig2}
\end{figure}
We first briefly introduce the definition of Fresnel zone. The Fresnel zone, supported by Huygens principle, elucidates the relation between geometry and propagation of wave. As shown in Fig.~\ref{fig2}, it describes the interference of the reflected wave regarding the direct link, where the $i$-th Fresnel zone is in the shape of ellipsoid defined as the paths with $i\pi$ propagation phase difference with respect to the shortest LoS link and the locations of BS and UE are the focal points of these ellipsoids\cite{RIS_fresnel}. The parameters and phase distributions of Fresnel zones are mainly determined by the location information and wavelength. Generally, in the far-field region, the Fresnel ellipsoid is narrow like a pencil, resulting in planar wave with nearly linear phase distributions. While in the near-field region, the phase distribution of Fresnel ellipsoid contains abundant curvature information, as shown in Fig.~\ref{fig2}, which helps to intuitively unveil the impact of locations on signal reception. Also, we denote by $\mathcal{D}$ and $\mathcal{I}$ regions the 'distinct region' near the parametric curve and the remaining 'indistinct region'. {The functions of such regions will be subsequently introduced.}

The relation between location and Fresnel ellipsoid can be constructed by formulating the ellipsoid equation. As shown in Fig.~\ref{fig2}, the diagram of the first and second Fresnel zones among the $k$-th RIS, the $n_t$-th BS antenna, and UE is depicted where $\boldsymbol{\rm v}=\left(v_{x},v_{y},v_{z}\right) \in \mathbb{C}^{1\times 3}$ denotes the arbitrary point at the 3D Cartesian coordinate system, $d^{TU}_{n_t}$, $d^{TR_k}_{n_t,\boldsymbol{\rm v}}$, and $d^{UR_k}_{\boldsymbol{\rm v}}$ denote the distance from the $n_t$-th BS antenna to UE, the distance from $n_t$-th BS antenna to the point $\boldsymbol{\rm v}$, and the distance from UE to the point $\boldsymbol{\rm v}$, respectively, which can be calculated by 
\begin{equation}
    \begin{cases}
          d^{TR_k}_{n_t,\boldsymbol{\rm v}}=\left\lVert \boldsymbol{\rm  t}_{n_t}-{\boldsymbol{\rm  v}} \right\rVert ; \\
          
          d^{UR_k}_{\boldsymbol{\rm v}}=\left\lVert \boldsymbol{\rm  u}-{\boldsymbol{\rm  v}}\right\rVert ; \\
          
          d^{TU}_{n_t}=\left\lVert {\boldsymbol{\rm  t}}_{n_t}-{\boldsymbol{\rm  u}}\right\rVert .
    \end{cases}
\end{equation}
Accordingly, the locus of point $\boldsymbol{\rm v}$ satisfying
\begin{align}
 d^{TR_k}_{n_t,\boldsymbol{\rm v}}+d^{UR_k}_{\boldsymbol{\rm v}}=d^{TU}_{n_t}+\frac{i\lambda}{2} \label{dif}
\end{align}
determines the $i$-th Fresnel ellipsoid. Let $d^{TU}_{n_t,1}$ and $d^{TU}_{n_t,1}$ denote the projection of $d^{UR_k}_{\boldsymbol{\rm v}}$ and $d^{TR_k}_{n_t,\boldsymbol{\rm v}}$ on the BS-UE link, respectively, which can be written as 
\begin{equation}
    \begin{cases}
        d^{TU}_{n_t,1}=\left({\boldsymbol{\rm  v}}-\boldsymbol{\rm  u}\right) \frac{\left({\boldsymbol{\rm  t}}_{n_t}-{\boldsymbol{\rm  u}}\right)^{\rm H}} {\left\lVert {\boldsymbol{\rm  t}}_{n_t}-{\boldsymbol{\rm  u}}\right\rVert };\\\vspace{0.2 cm}
        d^{TU}_{n_t,2}=\left(\boldsymbol{\rm  t}_{n_t}-{\boldsymbol{\rm  v}} \right) \frac{\left({\boldsymbol{\rm  t}}_{n_t}-{\boldsymbol{\rm  u}}\right)^{\rm H}} {\left\lVert {\boldsymbol{\rm  t}}_{n_t}-{\boldsymbol{\rm  u}}\right\rVert };\\\vspace{0.1 cm}
        d^{TU}_{n_t}~=d^{TU}_{n_t,1}+d^{TU}_{n_t,2}. 
    \end{cases}\label{distance}
\end{equation}
We denote by $r_i$ the Fresnel radius of the $i$-th Fresnel zone, which describes the radius of the section perpendicular to the semi-major axis\cite{RIS_fresnel}. Then, based on trigonometric relation, Eq.~\eqref{dif} can be rewritten as follows:
\begin{align}
    \sqrt{ \left(d^{TU}_{n_t,1}\right) ^2+r_i^2}+\sqrt{\left(d^{TU}_{n_t,2}\right) ^2+r_i^2}=d^{TU}_{n_t}+\frac{i\lambda}{2}. \label{diff2}
\end{align}
According to Fresnel approximation when $d^{TU}_{n_t,1},~d^{TU}_{n_t,2}>> \lambda$, Eq.~\eqref{diff2} can be rewritten as 
\begin{equation}
\begin{aligned}
    &~~~~~~~~\sqrt{ \left(d^{TU}_{n_t,1}\right) ^2+r_i^2}+\sqrt{\left(d^{TU}_{n_t,2}\right) ^2+r_i^2}\\&\approx
    d^{TU}_{n_t,1}+\frac{r_i^2}{2d^{TU}_{n_t,1}}+d^{TU}_{n_t,2}+\frac{r_i^2}{2d^{TU}_{n_t,2}}=d^{TU}_{n_t}+\frac{i\lambda}{2}.\label{diff3}
\end{aligned}    
\end{equation}
Uniting Eqs.~\eqref{distance} and \eqref{diff3}, the expression of Fresnel radius can be given as follows:
\begin{align}
    r_i = \sqrt{\frac{i\lambda  d^{TU}_{n_t,1} d^{TU}_{n_t,2}}{d^{TU}_{n_t}}}. \label{radius}
\end{align}
With Eq.~\eqref{radius}, the parameters of Fresnel ellipsoids by considering the special case can be obtained. We denote by $\rho^{a}_{n_t,i}$, $\rho^{b}_{n_t,i}$, and $\rho^{c}_{n_t,i}$ the lengths of semi-major axis, the semi-middle axis, and the semi-minor axis of the $i$-th Fresnel ellipsoid between the $n_t$-th BS antenna and UE, respectively. Hence, $\rho^{a}_{n_t,i}$ can be obtained from Eq.~\eqref{dif} by setting the point $\boldsymbol{\rm v}$ at left vertex of the ellipse, which can be expressed as $\rho^{a}_{n_t,i}=d^{TU}_{n_t}/2+i\lambda/{4}$. Also, the semi-middle axis $\rho^{b}_{n_t,i}$ can be obtained as the length of Fresnel radius by setting the point $\boldsymbol{\rm v}$ at top vertex of the ellipse, which can be expressed as $\rho^{b}_{n_t,i}={\sqrt{i\lambda d^{TU}_{n_t}}}/{2}$. The Fresnel ellipsoid has the rotational symmetry property with respect to the semi-major axis, leading to $\rho^{c}_{n_t,i}=\rho^{b}_{n_t,i}$. So far, the derived parameters with respect to Fresnel ellipsoids is sufficient to give its expressions. The key points are the parametric curves formulated by the intersection lines between RIS and Fresnel zones, which reveals the phase distribution of incident signal on RIS. 

\subsection{Extension to Fractional Fresnel Zone}
The Fresnel zone model gives the 1-bit case for phase distributions of the reflected paths, which only can be referred to configure RIS elements with 1-bit phase resolution. Hence, to derive a general result in consideration of $D$-bit elements, we extend the traditional Fresnel zone model with only integer index to the fractional Fresnel zone model to make it compatible with general phase resolution case. 

{In specific, considering $D$-bit phase resolution, the locus of point $\boldsymbol{\rm v}$ in fractional Fresnel zone model should satisfy 
\begin{align}
 d^{TR_k}_{n_t,\boldsymbol{\rm v}}+d^{UR_k}_{\boldsymbol{\rm v}}=d^{TU}_{n_t}+j\lambda, \label{dif1}
\end{align}
where $j=i/{2^{D}}$ denotes the new index of fractional Fresnel zone. When $D=1$, the fractional Fresnel zone is reduced to traditional Fresnel zone.} Accordingly, by substituting $j$ to traditional semi-axes of the Fresnel zone, the semi-axes of fractional Fresnel zone can be modified as follows:
\begin{equation}
   \begin{cases}
\rho_{n_t,j}^{a}=\frac{1}{4}\left({2d^{TU}_{n_t}}+{j\lambda}\right) ; \vspace{0.1 cm}\\
\rho_{n_t,j}^{b}=\frac{1}{2}\sqrt{j\lambda d^{TU}_{n_t}};\vspace{0.1 cm}\\
\rho_{n_t,j}^{c}=\frac{1}{2}\sqrt{j\lambda d^{TU}_{n_t}}, \label{modified_axes}
   \end{cases}
\end{equation}
where $\rho^{a}_{n_t,j}$, $\rho^{b}_{n_t,j}$, and $\rho^{c}_{n_t,j}$ the semi-major axis, the semi-middle axis, and the semi-minor axis of the $j$-th fractional Fresnel ellipsoid between the $n_t$-th BS antenna and UE, respectively. The physical meaning of such parameters is that, compared with traditional Fresnel zone model, the previous Fresnel zone is further subdivided to be compatible with RIS in high phase resolution case. Based on the modified parameters in Eq.~\eqref{modified_axes}, the parametric equation of fractional Fresnel ellipsoids can be constructed.

\subsection{Interaction between RIS and Fresnel Ellipsoid}
In Section III-A\&B, the mapping from locations to phase distributions is preliminarily revealed by the Fresnel diffraction theory, but uniting the Fresnel zone and proposed RIS model remains two urgent issues to solve 1) the element-wise selection criterion of determining the state of RIS elements; 2) what's the impact of phase deviations induced by discrete phase shift constraint on the selection criteria. 

\begin{figure}[t]
    \centerline{\includegraphics[width=8.5 cm]{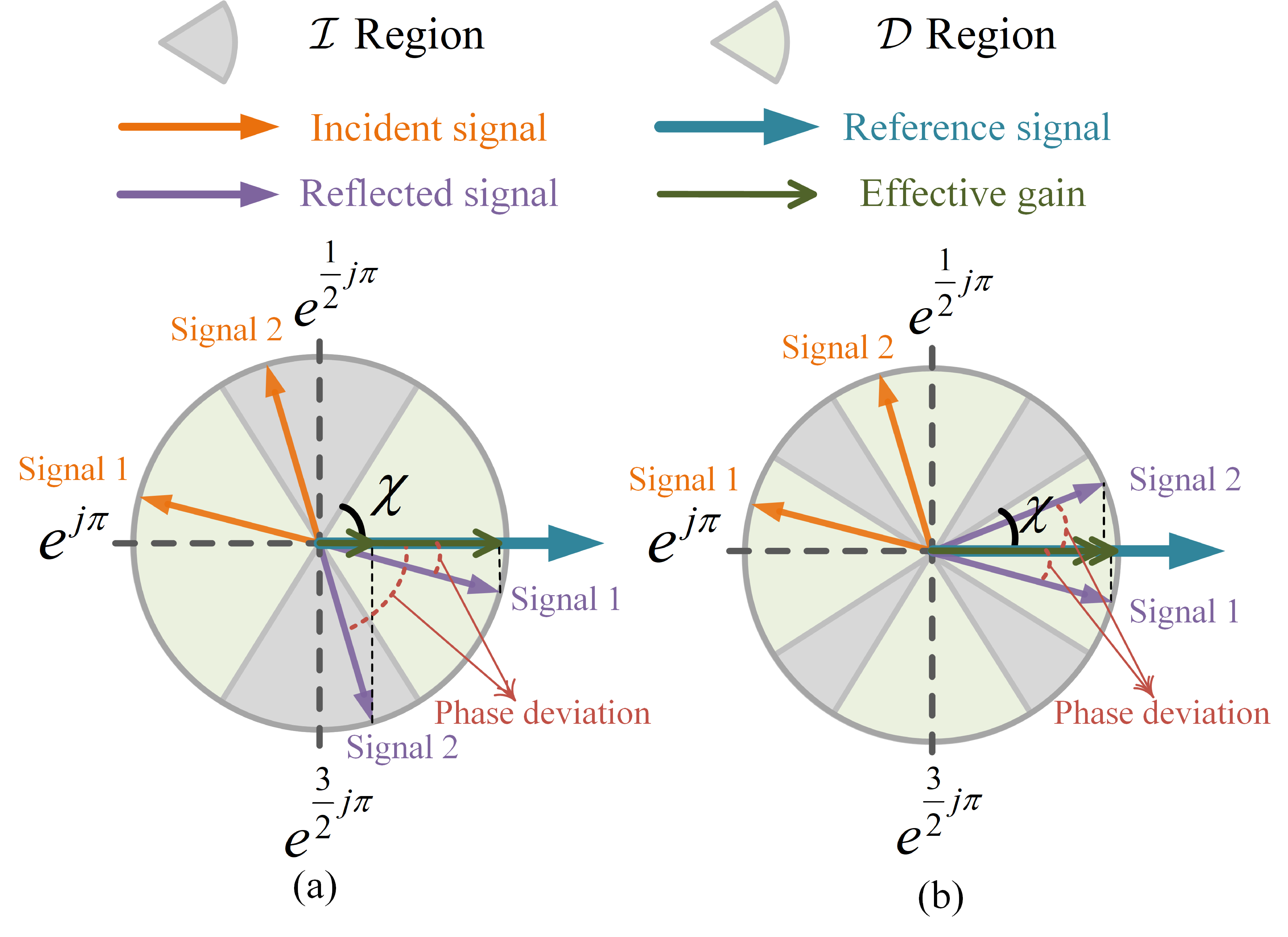}}
        \caption{{The diagrams of configuration criterion between reflective elements and absorptive elements: (a) the phase resolution of reflective element $D=1$; (b) the phase resolution of reflective element $D=2$.}}
        \label{fig3}
\end{figure}

As shown in Fig.~\ref{fig3}, the diagrams of element-wise selection criteria for reflective elements and absorptive elements under $D=1,2$ are given. {Therein, the reference signal denotes the signal transmitted by direct link, whose power is typically far greater than that of signals transmitted by reflected paths in RIS-assisted system}\cite{rank-one,rank-one2}. Hence, the key point is aligning the reflected path of RIS with the direct link to enhance the composed channel. {Nevertheless, due to the discrete phase shift constraint in practical scenarios, the signals reflected by RISs cannot perfectly align with the reference signal, inducing inevitable phase deviations. The phase deviation of RIS-associated path determines the final effective gain, expressed by the projection of reflected channel vector on the direct channel vector. The effective gain provided by reflected signals with large phase deviation is seriously limited, resulting in energy waste. More importantly, the inevitable localization error and mobility of UE lead to erroneous judgement of reflected signals, providing destructive effects on reference signal. Thereby, we define the $\mathcal{I}$ region set to gather such susceptible elements to reduce energy wasting. On the contrary, the elements with small phase deviations are allocated to the $\mathcal{D}$ region. To distinguish $\mathcal{I}$ and $\mathcal{D}$ regions by a specific parameter, we denote by $\chi$ half angle of the $\mathcal{D}$ region satisfying $\chi\in \left[0, \pi/2^D\right]$, which actually determines the element-wise selection criterion. To further elucidate the principle, we exemplify with 1-bit and 2-bit phase resolution cases to explicate its inherent mechanism of the previously defined $\mathcal{D}$ and $\mathcal{I}$ regions in the following.}

{\emph{1-bit case:} As shown in Fig.~\ref{fig3} (a), the phase shifts of reflective elements can be chosen from two cases, e.g., $\left\{0, \pi\right\} $. The incident signals 1 and 2 are located at the $\mathcal{D}$ and $\mathcal{I}$ regions, respectively. We denote by effective reflected gain the projection of reflected signal with respect to the reference signal. For reflection enhancement design, with the optimal phase shift, reflected signal 1 with $\pi$ phase shift can align well with the reference signal to provide desired effective reflected gain whereas reflected signal 2 can only provide limited gain. Meanwhile, all incident signals provide the same gain as there is no reference signal to align in EH design. With the same strength, it can be concluded that the signal 2 is more conducive to enhancing the received power at UE, implying that the elements in $\mathcal{D}$ region show higher priority to provide reflected gain. Therefore, the elements in $\mathcal{D}$ and $\mathcal{I}$ regions should be configured into reflective and absorptive modes, respectively, to achieve better energy efficiency. }

{\emph{2-bit case:} As shown in Fig.~\ref{fig3} (b), there are 4 phase shifts to configure reflective elements when $D=2$. As phase resolution increases, the previous $\mathcal{I}$ region gradually splits into smaller pieces. With the same incident signals 1 and 2, the reflected signal 1 behaves the same by introducing $\pi$ phase shift and still provides desired reflected gain. Differently, the incident signal 2 in previous $\mathcal{I}$ region can also provide desired reflected gain for the reference signal by introducing $\pi/2$ phase shift. It is noteworthy that higher phase resolution will not determine the proportion of the two regions, but the largest phase deviation caused by discrete phase shift constraint. Therefore, with larger phase resolution, the effective reflected gains of different elements gradually approach the similar level and are finally the same. According to the above analysis, the following \emph{Remark} is given.

\begin{remark}
    The reflected signals with small and large phase deviations are allocated to the $\mathcal{D}$ and $\mathcal{I}$ regions, respectively. {The number of partitions of two regions is determined by the phase resolution and is equal to $2^D$.} As the phase resolution increases, the otherness of different elements in the two regions can be gradually remedied by more fine-grained phase shifts and finally disappears when phase shift is perfect. Under such condition, elements can be arbitrarily selected to achieve the same optimal performance. Therefore, the performance gain of the proposed element-wise criterion gradually decreases as phase shift resolution increases and finally approaches 1 with the ideal phase shift. 
\end{remark}
}

\subsection{Parametric Curve}
The parametric curves can be derived by solving joint equations constructed by Fresnel ellipsoids and RISs. Without loss of generality, we stipulate an established principle of Cartesian coordinate system to simplify the derivation. For each RIS, the projection point of UE's position on RIS plane is selected as the coordinate with RIS array being in $xz$-plane, which can be easily addressed by coordinate transformation based on the absolute locations of RIS and transceiver. Under this principle, we denote by $\hat{{\boldsymbol{\rm  t}}}_{n_t}=\left[\hat{t}_{n_t,x},\hat{t}_{n_t,y},\hat{t}_{n_t,z}\right] $, $\hat{\boldsymbol{\rm  r}}^k_{n_r}=\left[\hat{r}^k_{n_r,x},\hat{r}^k_{n_r,y},\hat{r}^k_{n_r,z}\right] $, and $\hat{\boldsymbol{\rm u}}=\left[\hat{u}_{x},\hat{u}_{y},\hat{u}_{z}\right] $ the new coordinates of the $n_t$-th BS antenna, the $n_r$-th element of the $k$-th RIS, and UE, respectively. Hence, the midpoint of the link from the $n_t$-th BS antenna and UE, denoted by $\boldsymbol{\rm c}_{n_t}$, can be expressed as follows:
\begin{equation}
\begin{aligned}
    &\boldsymbol{\rm c}_{n_t}=\frac{\hat{{\boldsymbol{\rm  t}}}_{n_t}+\hat{\boldsymbol{\rm u}} +\boldsymbol{\rm  n}_l}{2}=\\
    &\left[\frac{\hat{t}_{n_t,x}+\hat{u}_{x}+n_{l,x}}{2},\frac{\hat{t}_{n_t,y}+\hat{u}_{y}+n_{l,y}}{2},\frac{\hat{t}_{n_t,z}+\hat{u}_{z}+n_{l,z}}{2}\right],
\end{aligned}    
\end{equation}
{which also represents the center of corresponding Fresnel ellipsoids.} Based on the parameters in Eq.~\eqref{modified_axes}, we first consider the standard form of Fresnel ellipsoid equations. The motion trail of $\boldsymbol{\rm v}$ with respect to the $i$-th Fresnel ellipsoid between the $n_t$-th BS antenna and UE satisfies
\begin{align}
    \frac{{v_x}^2}{{\rho^{a}_{n_t,i}}^2}+\frac{{v_y}^2}{{\rho^{b}_{n_t,i}}^2}+\frac{{v_z}^2}{{\rho^{c}_{n_t,i}}^2}=1,
    \label{ellipsoid equation}
\end{align}
where the semi-major axis coincides with the $x$-axis. To derive intuitive results, the standard form in Eq.~\eqref{ellipsoid equation} is transformed from Cartesian coordinate system into polar coordinate system as follows:
\begin{align}
    \boldsymbol{\rm v}^T=
    \begin{bmatrix}
        \rho^{a}_{n_t,i} \sin \vartheta \cos \varphi\\ \rho^{b}_{n_t,i}\sin \vartheta \sin \varphi\\\rho^{c}_{n_t,i}\cos \vartheta 
    \end{bmatrix}. \label{coordinate}
\end{align}
It is noteworthy that any equation of ellipsoid can be derived from Eq.~\eqref{coordinate} by translation and rotation. To derive the actual distributions of Fresnel ellipsoids in consideration of BS and UE, we denote by $\alpha$ and $\beta$ the azimuth and elevation angles of the vector $\hat{\boldsymbol{\rm u}}-\hat{{\boldsymbol{\rm  t}}}_{n_t}$, which can be expressed as by
\begin{equation}
\begin{aligned}
    \alpha&={\rm arctan}\left(\frac{\left(\hat{\boldsymbol{\rm u}}+\boldsymbol{\rm  n}_l-\hat{{\boldsymbol{\rm  t}}}_{n_t}\right){\boldsymbol{\rm i}_y}^T }{\left(\hat{\boldsymbol{\rm u}}+\boldsymbol{\rm  n}_l-\hat{{\boldsymbol{\rm  t}}}_{n_t}\right){\boldsymbol{\rm i}_z}^T }\right) \\
    &\approx{\rm arctan}\left(\frac{\left(\hat{\boldsymbol{\rm u}}-\hat{{\boldsymbol{\rm  t}}}_{n_t}\right){\boldsymbol{\rm i}_y}^T }{\left(\hat{\boldsymbol{\rm u}}-\hat{{\boldsymbol{\rm  t}}}_{n_t}\right){\boldsymbol{\rm i}_z}^T }\right) 
\end{aligned}    
\end{equation}
and 
\begin{equation}
\begin{aligned}
    \beta&={\rm arctan}\left(\frac{\left(\hat{\boldsymbol{\rm u}}+\boldsymbol{\rm  n}_l-\hat{{\boldsymbol{\rm  t}}}_{n_t}\right){\boldsymbol{\rm i}_z}^T }{\left\lVert \hat{\boldsymbol{\rm u}}+\boldsymbol{\rm  n}_l-\hat{{\boldsymbol{\rm  t}}}_{n_t}\right\rVert  }\right) \\
    &\approx{\rm arctan}\left(\frac{\left(\hat{\boldsymbol{\rm u}}-\hat{{\boldsymbol{\rm  t}}}_{n_t}\right){\boldsymbol{\rm i}_z}^T }{\left\lVert \hat{\boldsymbol{\rm u}}-\hat{{\boldsymbol{\rm  t}}}_{n_t}\right\rVert  }\right) ,
\end{aligned}    
\end{equation}
respectively, where $\boldsymbol{\rm i}_x$, $\boldsymbol{\rm i}_x$, and $\boldsymbol{\rm i}_x$ denote the unit vector in the directions of the $x$-axis, $y$-axis, and $z$-axis, respectively, and the approximations hold as $\left\lVert \boldsymbol{\rm  n}_l\right\rVert \ll \left\lVert \hat{\boldsymbol{\rm u}}-\hat{{\boldsymbol{\rm  t}}}_{n_t}\right\rVert $ holds for typical communication distances. Thereby, the parametric curve of the $j$-th fractional Fresnel ellipsoid between the $n_t$-th BS antenna and UE, denoted by $\boldsymbol{\rm v}_{n_t,j}$, can be written as 
\begin{align}
    \boldsymbol{\rm v}_{n_t,j}^T={\boldsymbol{\rm c}_{n_t}}^T+\boldsymbol{\rm R}\left(\alpha,\beta\right) 
    \begin{bmatrix}
        \rho_{n_t,j}^{a} \sin \vartheta \cos \varphi\\ \rho_{n_t,j}^{b}\sin \vartheta \sin \varphi\\\rho_{n_t,j}^{c}\cos \vartheta 
    \end{bmatrix}, \label{curve}
\end{align}
where $\boldsymbol{\rm R}\left(\alpha,\beta\right)$ denote the rotation matrix. Due to the rotational symmetry of the Fresnel ellipsoid corresponding to $x$-axis, rotations in two directions including $y$-axis and $z$-axis are sufficient to characterize the motion of an arbitrary ellipsoid, rather than in three directions. $\boldsymbol{\rm R}\left(\alpha,\beta\right)$ can be expressed in Eq.~\eqref{rotation}.
\begin{figure*}[!tb]
	{\noindent} 
    \begin{equation}
        \begin{aligned}
            \boldsymbol{\rm R}\left(\alpha,\beta\right)=\boldsymbol{\rm R}_z\left(\alpha\right)\boldsymbol{\rm R}_y\left(\beta\right)
            =
            \begin{bmatrix}
                \cos \alpha &-\sin \alpha & 0 \\
                \sin \alpha &\cos \alpha& 0\\
                0& 0 & 1 
            \end{bmatrix}
            \begin{bmatrix}
                \cos \beta &-\sin \beta & 0 \\
                  0& 0 & 1 \\
                \sin \beta &\cos \beta& 0
            \end{bmatrix}
            =
            \begin{bmatrix}
                \cos \alpha \cos \beta&-\cos \alpha \sin \beta& -\sin \alpha \\
                \sin \alpha \cos \beta&-\sin \alpha\sin\beta& \cos \alpha\\
                \sin \beta& \cos \beta & 0
            \end{bmatrix}
        \end{aligned}
        \label{rotation}
    \end{equation}
    \rule[-7pt]{18.05cm}{0.05em}	
\end{figure*}
Therefore, the parametric equation in Eq.~\eqref{curve} can be rewritten as Eq.~\eqref{curve2}.
\begin{figure*}[!tb]
	{\noindent} 
    \begin{equation}
        \begin{aligned}
            \boldsymbol{\rm v}_{n_t,j}^T=
            \begin{bmatrix}
                \rho_{n_t,j}^{a} \sin \vartheta \cos \varphi  \cos \alpha \cos \beta -\rho_{n_t,j}^{b}\sin \vartheta \sin \varphi\cos \alpha \sin \beta-\rho_{n_t,j}^{c}\cos \vartheta \sin \alpha+\frac{\hat{t}_{n_t,x}+\hat{u}_{x}+n_{l,x}}{2}\\
                \rho_{n_t,j}^{a} \sin \vartheta \cos \varphi\sin \alpha \cos \beta-\rho_{n_t,j}^{b}\sin \vartheta \sin \varphi\sin \alpha\sin\beta+\rho_{n_t,j}^{c}\cos \vartheta \cos \alpha+\frac{\hat{t}_{n_t,y}+\hat{u}_{y}+n_{l,y}}{2}\\
                \rho_{n_t,j}^{a} \sin \vartheta \cos \varphi\sin \beta+ \rho_{n_t,j}^{b}\sin \vartheta \sin \varphi \cos \beta+\frac{\hat{t}_{n_t,z}+\hat{u}_{z}+n_{l,z}}{2}
            \end{bmatrix} 
        \end{aligned}
        \label{curve2}
    \end{equation}
    \rule[-7pt]{18.05cm}{0.05em}	
\end{figure*}
The parametric curve in the RIS plane locates at the $xy$-plane satisfying
\begin{equation}
    \begin{aligned}
        \rho_{n_t,j}^{a} \sin \vartheta \cos \varphi\sin \beta+& \rho_{n_t,j}^{b}\sin \vartheta \sin \varphi \cos \beta+~~~~~~~~~~~~~~\\
        &~~~~~~~~~~~~~\frac{\hat{t}_{n_t,z}+\hat{u}_{z}+n_{l,z}}{2}=0, \label{zero}
    \end{aligned}
\end{equation}
which can be rewritten as 
\begin{equation}
    \begin{aligned}
        \sin \vartheta \sin \varphi=-\frac{2\rho_{n_t,j}^{a}\sin \vartheta \cos \varphi \sin \beta+\hat{t}_{n_t,z}+\hat{u}_{z}+n_{l,z}}{2 \rho_{n_t,j}^{b} \cos \beta}. \label{equation1}
    \end{aligned}
\end{equation}
By substituting Eqs.~\eqref{zero} and \eqref{equation1} into Eq.~\eqref{curve2}, Eq.~\eqref{curve3} on the top of this page can be derived.
\begin{figure*}[!tb]
	{\noindent} 
    \begin{equation}
        \begin{aligned}
            \boldsymbol{\rm v}_{n_t,j}^T=
            \begin{bmatrix}
                \rho_{n_t,j}^{a} \left(\cos\alpha \cos\beta +\cos \alpha \sin \beta \tan \beta\right)\sin \vartheta \cos\varphi -\rho_{n_t,j}^{c} \sin \alpha \cos \vartheta+\frac{\cos\alpha\tan\beta\left(\hat{t}_{n_t,z}+\hat{u}_{z}+n_{l,z}\right) }{2} + \frac{\hat{t}_{n_t,x}+\hat{u}_{x}+n_{l,x}}{2}\\
                \rho_{n_t,j}^{a} \left(\sin\alpha \cos\beta +\sin \alpha \sin \beta \tan \beta\right)\sin \vartheta \cos\varphi +\rho_{n_t,j}^{c} \cos \alpha \cos \vartheta+ \frac{\sin\alpha\tan\beta\left(\hat{t}_{n_t,z}+\hat{u}_{z}+n_{l,z}\right) }{2} + \frac{\hat{t}_{n_t,y}+\hat{u}_{y}+n_{l,y}}{2}\\
                0
            \end{bmatrix} =\\
            \begin{bmatrix}
                \rho_{n_t,j}^{a} \left(\cos\alpha \cos\beta +\cos \alpha \sin \beta \tan \beta\right)\sin \vartheta \cos\varphi -\rho_{n_t,j}^{c} \sin \alpha \cos \vartheta+\frac{\hat{t}_{n_t,x}+\hat{u}_{x}+\cos\alpha\tan\beta\left(\hat{t}_{n_t,z}+\hat{u}_{z}\right) }{2} + \underbrace{\frac{\cos\alpha\tan\beta n_{l,z}+n_{l,x}}{2}}_{\rm noise~item}\vspace{0.3 cm}\\ 
                \rho_{n_t,j}^{a} \left(\sin\alpha \cos\beta +\sin \alpha \sin \beta \tan \beta\right)\sin \vartheta \cos\varphi +\rho_{n_t,j}^{c} \cos \alpha \cos \vartheta+ \frac{\hat{t}_{n_t,y}+\hat{u}_{y}+\sin\alpha\tan\beta\left(\hat{t}_{n_t,z}+\hat{u}_{z}\right) }{2} + \underbrace{\frac{\sin\alpha\tan\beta n_{l,z}+n_{l,y}}{2}}_{\rm noise~item}\\
                0
            \end{bmatrix}
        \end{aligned}
        \label{curve3}
    \end{equation}
    \rule[-7pt]{18.05cm}{0.05em}	
\end{figure*}
In the presence of Gaussian localization errors, it can be observed that the parametric curve in Eq.~\eqref{curve3}, which determines phase distributions of Fresnel zones, is also impacted by Gaussian noise items. Therein, the noise items in the $x$-axis and $y$-axis directions are amplified by the coefficients $\cos\alpha\tan\beta$ and $\sin\alpha\tan\beta$, respectively. {Additionally, such results can be also extended to the case considering the inevitable UE's mobility. If the parameters of motion, e.g., velocity and move direction, can be well sensed, the predicted location can be derived correspondingly. Then, the proposed parametric curve can well predict the phase distributions for the next time, followed by predictive beamforming to offset the induced outdated error. On the contrary, the error induced by mobility can be classified as location error with the same impact on parametric curve as concluded above, exhibiting favorable compatibility for UE's mobility. Hence, the following \emph{Remark} can be given.}

{
\begin{remark}
    All the error components in Eq.~\eqref{curve3} are Gaussian distributions. Additionally, the error items achieve the minimum value with the minimum impact on performance when $\tan \beta=0$, i.e., the BS-UE link is parallel to the RIS plane. Nevertheless, due to the property of $\tan \beta$ function, small alignment errors are tolerable in practical scenarios. In the scenario considering the UE's mobility, the outdated error induced by outdated location can be offset by predictive location or classified as location error with the same impact on parametric curve.
\end{remark}
}

So far, the phase distributions of Fresnel zones for common cases are derived, with which the joint absorptive and reflective coefficients of RIS elements can be designed.

{
\section{Distributed Transmission Design}
In this section, the distributed transmission scheme is designed, embodying the frame structure, the distributed design of initialization, and distributed beamforming, followed by asymptotic results to quantify the achievable gain. 

\begin{figure}[t]
    \centerline{\includegraphics[width=8.7cm]{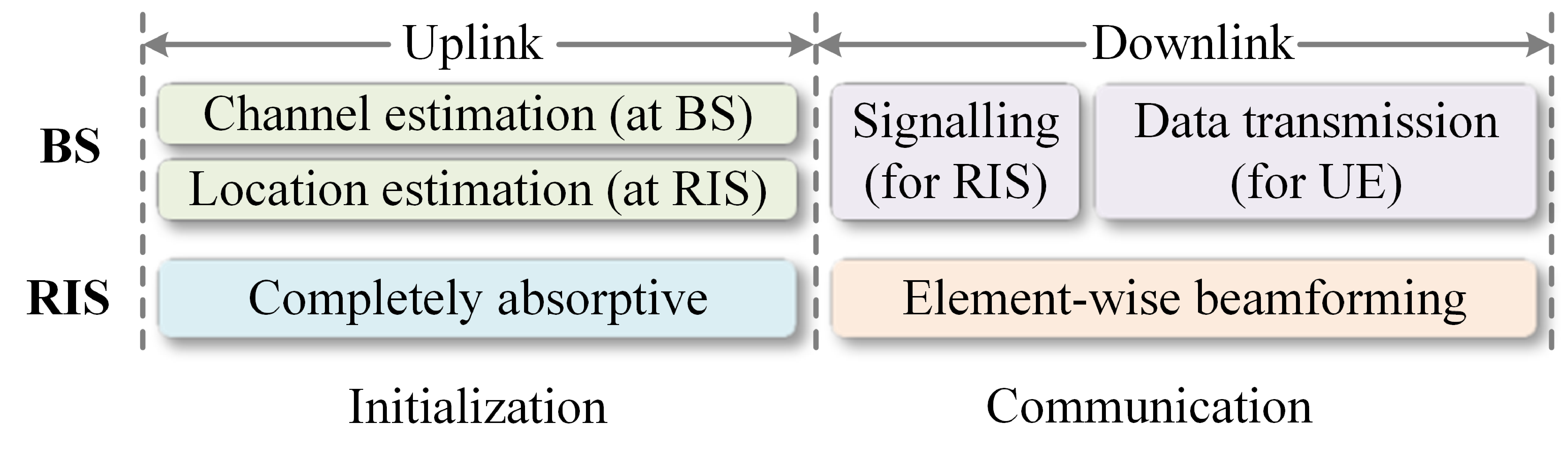}}
        \caption{The proposed frame structure.}
        \label{frame}
\end{figure}
We commence by designing the frame structure. As shown in Fig~\ref{frame}, under the time division protocol, each slot contains uplink and downlink stages. In the uplink, the RISs are configured to completely absorptive elements, which is referred to as the initialization stage. Within the initialization stage, the BS and RIS separately operate CE and location estimation (LE), respectively, to unload the complexity induced by centralized control. Based on the estimated results, the precoding design can be determined as the BS only derive the composite channel information in the absence of RIS-associated channel. Then, in the downlink, the BS firstly transmits the precoding results to the RIS controller through the dedicated signalling channel. With the precoding results and UE's location information, the element-wise beamforming at the RIS can be designed to align with the determined BS beamforming, called the phase of determine-then-align. 

\subsection{Initialization stage}
The initialization stage mainly comprises two separate estimation targets at BS and RIS, respectively, with all RIS elements being absorptive. Such design targets at a lightweight sense to not only reduce the complexity of joint estimation at the BS but also leverage the unique superiority for UE localization at the RIS, while providing initial power supply to subsequent operations.

In specific, in the absence of RIS-associated channel, the CE requirement in this paper is reduced to a simple multiple-input-single-output (MISO) CE problem, which has been widely investigated by existing works comprising both far-field and near-field cases under LoS/NLoS conditions\cite{CE6,CE3,CE4}. For example, typical far-field CE techniques has been fully summarized in \cite{CE6}. While for more complicated near-field channel, the authors in \cite{CE3} leveraged the polar-domain sparsity of the near-field channel, where an on-grid polar-domain simultaneous orthogonal matching pursuit algorithm is proposed to efficiently estimate the near-field channel in extremely large-scale multiple-input-multiple-output (XL-MIMO) systems. A mixed LoS/NLoS near-field XL-MIMO channel model is proposed to match the practical near-field XL-MIMO scenario in \cite{CE4}. Such efficient CE schemes can effectively address the demand of CE in the considered scenario when RIS is completely absorptive. Without RIS-associated channel, the estimated channel, denoted by $\boldsymbol{\rm \hat{f}}\in \mathbb{C}^{N_T\times 1}$, can be derived without much estimation overhead compared with traditional CSI-driven RIS-assisted schemes.

Also, to achieve a more lightweight RIS configuration, we release the complicated joint LE and CE tasks at BS in the proposed scheme. Instead, as the RIS configuration design completely depends on the UE's location, the LE is separately accomplished by the RIS controller. Therein, as the RIS is typically located near the UE side with fine-grained localization potential, the location of UE can be inferred using a single RIS from the curvature of incident wave induced by spherical wavefront in the near-field region\cite{Spherical_Localization}, which outperforms the traditional far-field trilateration. Hence, through configuring the absorptive coefficients of elements according to specific potential location, UE's location can be well captured by using the power indicator, where the maximum received power corresponds to the true UE's location. In specific, for each test location $\hat{\boldsymbol{\rm u}}$, we denote by $\boldsymbol{\rm d}_k \in \mathbb{C}^{1\times N_R}$ the distance from the $k$-th RIS to $\hat{\boldsymbol{\rm u}}$, which can be expressed by 
\begin{equation}
    \boldsymbol{\rm d}_k=\left[\left\lVert \boldsymbol{\rm  r}^k_{1}-\boldsymbol{\rm  u}\right\rVert, \cdots, \left\lVert \boldsymbol{\rm  r}^k_{N_R}-\boldsymbol{\rm  u}\right\rVert\right].
\end{equation}
{Then, the $n_r$-th element of the switch array is configured to 
\begin{equation}
   \boldsymbol{\Lambda}_k(n_r)= \varsigma_{k,n_r}=
    \begin{cases}
        0, ~~\text{if}~~ \frac{\boldsymbol{\rm d}_k(n_r)}{\lambda} \in \left(\frac{\left(4n-1\right)\pi }{2}, \frac{\left(4n+1\right)\pi }{2}\right];\\
        \pi,~~\text{if}~~ \frac{\boldsymbol{\rm d}_k(n_r)}{\lambda} \in \left(\frac{\left(4n+1\right)\pi }{2}, \frac{\left(4n+3\right)\pi }{2}\right],
    \end{cases}
    \label{switch1}
\end{equation}
for $n=0,1,2,\cdots$ to maximize the received power if the UE is located at $\hat{\boldsymbol{\rm u}}$.} Hence, by varying $\boldsymbol{\rm u}$ within a set of test positions during the uplink, the UE's location can be estimated through the power indicator as follows:
\begin{equation}
   {\boldsymbol{\rm u}}= \max_{\hat{\boldsymbol{\rm u}}} \left\lVert \boldsymbol{\Lambda}_k\boldsymbol{\rm g}_{k}\right\rVert^2.
\end{equation}

Such procedure is similar to that of the far-field beam training. Nevertheless, the difference is that the distance information can be further captured thanks to the processing at electromagnetic and signal levels of RIS in the near-field region\cite{Spherical_Localization}.}

\subsection{Distributed Beamforming Design}
In the downlink, as opposed to traditional centralized beamforming schemes, we propose the distributed design with the step of determine-then-align, where BS determines the optimal precoding vector and then RIS can autonomously align the reflected signals. In specific, the BS adopts the maximum ratio transmission (MRT) satisfying $\boldsymbol{\rm w}={\boldsymbol{\rm \hat{f}}^H}/{\left \| \boldsymbol{\rm \hat{f}} \right \| }$ without considering the RIS-associated channel. Then, at the beginning of the downlink, the estimated channel parameter $\boldsymbol{\rm \hat{f}}$ and UE's location are transmitted from BS to the RIS controller through the dedicated control channel. Based on such priori information, the reflective and absorptive RIS coefficients can be independently designed by RIS controller, followed by the data transmission. {It is noteworthy that the LoS and NLoS components do not need to be separately estimated. The essential information at BS in the proposed frame is the composite channel including both LoS and NLoS components to operate MRT precoding.}

{With localization and CE results, the derived parametric curves characterize the phase distributions of reflected waves. By uniting the proposed element selection criteria with parametric curves, the joint design of reflective and absorptive elements can be determined.} According to the element selection criterion, the elements in $\mathcal{D}$ and $\mathcal{I}$ region should be configured to reflective and absorptive modes, respectively. The next step is to establish a parametric relation between the angle of the $\mathcal{D}$ region $\chi$ and the parametric curve in Eq.~\eqref{curve3}. In Fresnel zone model, the parametric curves are divided by phases. Based on the concept of differential, the pieces divided by phases can be very small, and thus the following equation can be given:
\begin{equation}
    \frac{2^D\chi}{\pi}=\frac{\tau_{n_t,j}}{\min \left\lVert \boldsymbol{\rm v}_{n_t,j+\frac{1}{2^{D}}}-\boldsymbol{\rm v}_{n_t,j}\right\rVert }, \label{proportion}
\end{equation}
where $\tau_{n_t,j}$ denotes the decision threshold for the $j$-th Fresnel zone. The physical significance of Eq.~\eqref{proportion} is that the proportion of the $\mathcal{D}$ region in phase domain is equal to that of in space domain. Hence, $\tau_{n_t,j}$ can be expressed as follows:
\begin{equation}
    \tau_{n_t,j}=\frac{{\min \left\lVert \boldsymbol{\rm v}_{n_t,j+\frac{1}{2^{D}}}-\boldsymbol{\rm v}_{n_t,j}\right\rVert }2^D\chi}{\pi}. \label{threshold}
\end{equation}
Recalling that in Section IV-A, the BS adopts the MRT, and thus the actual Fresnel zones between BS and UE should be revised according to the induced phases by the BS precoding. In specific, in each time slot, the phase offset induced by MRT, denoted by $\boldsymbol{\Psi }$, is $ \boldsymbol{\Psi }=\angle \boldsymbol{\rm \hat{f}}=\left [\Psi_1, \cdots,  \Psi_{N_T} \right ]^{\rm T}$ with $\boldsymbol{\Psi }\in \mathbb{C}^{N_T\times 1}$. To eliminate the induced phase offset, the simi-axes of fractional Fresnel ellipsoids should be modified as follows:
\begin{equation}
    \begin{cases}
 \hat{\rho}_{n_t,j}^{a}=\frac{1}{4}\left({2d^{TU}_{n_t}}+{j\lambda}-\frac{\Psi_{n_t} \lambda}{2\pi}\right) ; \vspace{0.1 cm}\\
 \hat{\rho}_{n_t,j}^{b}=\frac{1}{2}\sqrt{j\lambda d^{TU}_{n_t}};\vspace{0.1 cm}\\
 \hat{\rho}_{n_t,j}^{c}=\frac{1}{2}\sqrt{j\lambda d^{TU}_{n_t}}.
    \end{cases}
    \label{modified_rho}
\end{equation}
By substituting Eq.~\eqref{modified_rho} to Eq.~\eqref{curve3}, the modified parametric curve can be derived, denoted by $\boldsymbol{\rm \hat{v}}_{n_t,j}$. We denote by $\boldsymbol{\Phi}^{\rm r}_{k,n_t,j}$ the configuration matrix of the $k$-th RIS related to the $j$-th fractional Fresnel zone and the $n_t$-th BS antenna. Therefore, according to \emph{Remark 2}, the phase shift design is similar to the ML detection problem with Gaussian noise, resembling the minimum distance criterion for signal detection, which can be expressed by
\begin{equation}
    \boldsymbol{\Phi}^{\rm r}_{k,n_t,j}\left ( n_r \right ) =
    \begin{cases}
       1,\quad \text{if }\left \| \boldsymbol{\rm \hat{v}}_{n_t,j}- \hat{\boldsymbol{\rm  r}}^k_{n_r}\right \| \leq \tau_{n_t,j};\\
       0,\quad\quad \quad \quad \text{else}
    \end{cases}
\end{equation}
for $k=1,\cdots,K$ and $j=\frac{1}{2^{D}},\frac{2}{2^{D-1}},\cdots,J$, respectively, where $J$ denotes the maximum index of the fractional Fresnel zone. The spacing of adjacent Fresnel zones decreases as the index $j$ increases. When two Fresnel zones locate at the same RIS element, the incident waves from two Fresnel zones are coupled, making phase manipulation function of RIS element invalid. Hence, the upper bound $J$ is to avert such condition, which can be written as $\text{max}\left\{J\vert \left\| \boldsymbol{\rm \hat{v}}_{n_t,j+\frac{1}{2^{D}}}- \boldsymbol{\rm \hat{v}}_{n_t,j}\right \|-s_R>0\right\} $ and derived by numerical solution, where $s_R$ is the spacing of RIS elements typically with the value of $\lambda/10\sim \lambda/2$\cite{RIS_Path_Loss}. {To ensure the constructive interference condition among BS antennas, the following element selection strategy with respect to the $k$-th RIS and $j$-th fractional Fresnel zone, denoted by $\boldsymbol{\Phi}^{\rm r}_{k,j}$, should be considered:}
\begin{equation}
    \boldsymbol{\Phi}^{\rm r}_{k,j}=\underbrace{\boldsymbol{\Phi}^{\rm r}_{k,1,j}\odot \cdots \odot\boldsymbol{\Phi}^{\rm r}_{k,1,N_T}}_{\text{for}~~n_t=1,\cdots,N_T}. \label{product}
\end{equation}
Then, based on selection of RIS array, the final configuration matrix for the $k$-th RIS in reflective mode can be expressed as follows:
\begin{equation}
    \boldsymbol{\Phi}^{\rm r}_{k}=\sum_{j=\frac{1}{2^{D}}}^{J}\text{mod}\left ( j\pi,2\pi \right )\boldsymbol{\Phi}^{\rm r}_{k,j}.
    \label{reflective}
\end{equation}

To harness the full potential of EH module, apart from these reflective elements, all remaining elements will be configured in absorptive mode. Hence, $\boldsymbol{\Phi}^{\rm a}_k$ can be expressed as follows:
\begin{equation}
    \boldsymbol{\Phi}^{\rm a}_k=\varrho ^{\rm a}\lceil \frac{\boldsymbol{\Phi}^{\rm r}_{k}}{2} \rceil.
    \label{absorptive}
\end{equation}
For the absorptive elements, the switch array  $\boldsymbol{\Lambda}_k$ also should be designed to maximize the strength of harvest signals. We denote by $\boldsymbol{\rm K}_{k}=\text{diag}\left({\boldsymbol{\kappa}_k}\right)  \in \mathbb{C}^{N_R\times N_R}$ the wave-number matrix related to the $k$-th RIS, where $\boldsymbol{\rm\kappa}_k=\left[\kappa _{k,1},\cdots,\kappa _{k,N_R}\right] $. Based on the distance from the BS to the $k$-th RIS and the phase offset induced by MRT, $\boldsymbol{\rm\kappa}_k$ can be expressed as follows:
\begin{equation}
    \boldsymbol{\rm\kappa}_k =\frac{\boldsymbol{1}^{1\times N_T}\left(2\pi\boldsymbol{\rm D}_{k}+\boldsymbol{\Psi}\otimes  \boldsymbol{1}^{1\times N_R}\right) }{2\pi\lambda}, \label{wavenumber}
\end{equation}
where $\boldsymbol{\rm D}_{k}\in \mathbb{C}^{N_T\times N_R}$ denotes the distance matrix from the BS to the $k$-th RIS with the following expression:
\begin{equation}
    \boldsymbol{\rm D}_{k}=
    \begin{bmatrix}
        \left\lVert \boldsymbol{\rm  t}_{1}-\boldsymbol{\rm  r}^k_{1}\right\rVert& \cdots&\left\lVert \boldsymbol{\rm  t}_{1}-\boldsymbol{\rm  r}^k_{n_r}\right\rVert\\
        \vdots&\ddots &\vdots\\
        \left\lVert \boldsymbol{\rm  t}_{1}-\boldsymbol{\rm  r}^k_{n_t}\right\rVert& \cdots&\left\lVert \boldsymbol{\rm  t}_{n_t}-\boldsymbol{\rm  r}^k_{n_r}\right\rVert
      \end{bmatrix}.
\end{equation}
{Based on interference condition, the $n_r$-th diagonal element of $\boldsymbol{\Lambda}_k$ can be designed as follows:
\begin{equation}
   \boldsymbol{\Lambda}_k(n_r)=\varsigma_{k,n_r}=
    \begin{cases}
        0, ~~\text{if}~~ \boldsymbol{\rm\kappa}_k(n_r) \in \left(\frac{\left(4n-1\right)\pi }{2}, \frac{\left(4n+1\right)\pi }{2}\right];\\
        \pi,~~\text{if}~~ \boldsymbol{\rm\kappa}_k(n_r) \in \left(\frac{\left(4n+1\right)\pi }{2}, \frac{\left(4n+3\right)\pi }{2}\right],
    \end{cases}
    \label{switch}
\end{equation}
for $n=0,1,2,\cdots$.} As a result, based on Eqs.~\eqref{reflective}, \eqref{absorptive}, and \eqref{switch}, a suboptimal solution of $\boldsymbol{\Phi}^{\rm a}_k$, $\boldsymbol{\Phi}^{\rm r}_k$, $\boldsymbol{\Lambda}_k$ can be derived in closed form for the optimization problem \textbf{\textit{P}1} by constructing an interpretable model based on Fresnel diffraction theory. The detailed procedures are summarized in \textbf{Algorithm 1}.
\vspace*{0.4cm}
\begin{algorithm}[t]
    \caption{The location-aided distributed beamforming} 
    \hspace*{0.02in} 
   \begin{algorithmic}[1]    
    \STATE{\bf Initialization: $k=1$,~$n_r=1$,~$n_t=1$}
    \STATE {$j=\frac{1}{2^{D}}$;}     
    \WHILE{$k < K$}
    \WHILE{$j < J$}
    \WHILE{$n_t < N_T$}
    \WHILE{$n_r < N_R$}
    \IF {$\left\| \boldsymbol{\rm \hat{v}}_{n_t,j}- \hat{\boldsymbol{\rm  r}}^k_{n_r}\right \| \leq \tau_{n_t,j}$}
    \STATE {$\boldsymbol{\Phi}^{\rm r}_{k,n_t,j}\left ( n_r \right )\Leftarrow 1$}
    \ELSE
    \STATE {$\boldsymbol{\Phi}^{\rm r}_{k,n_t,j}\left ( n_r \right )\Leftarrow 0$}
    \ENDIF    
    \STATE $n_r =n_r+1$
    \ENDWHILE
    \STATE $n_t =n_t+1$
    \ENDWHILE
    \STATE $\boldsymbol{\Phi}^{\rm r}_{k,j}$ can be derived by Eq.~\eqref{product};
    \STATE $j =j+\frac{1}{2^{D}}$     
    \ENDWHILE
    \STATE $\boldsymbol{\Phi}^{\rm r}_{k}$ can be derived by Eq.~\eqref{reflective}; $\boldsymbol{\Phi}^{\rm a}_k$ can be derived by Eq.~\eqref{absorptive}; $\boldsymbol{\Lambda}_k$ can be derived by Eq.~\eqref{switch}; 
    \STATE $k =k+1$    
    \ENDWHILE
    \end{algorithmic}
\end{algorithm}

The computational complexity of the proposed scheme comprises searching the upper bound $J$ and designing the joint reflective and absorptive coefficients as well as the switch array. Therein, to derive $J$, the complexity with respect to the number of RIS elements and BS antenna is $\mathcal{O}\left(2^{D-1}J K N_R N_T \right)$. The complexity corresponding to reflective coefficients design is $\mathcal{O}\left(2^{D-1} J K N_R N_T \right)$. The complexity corresponding to reflective coefficients design is $\mathcal{O}\left(K N_R\right)$. The complexity corresponding to reflective coefficients design is $\mathcal{O}\left(K N_R N_T\right)$. Hence, the overall computational complexity is $\mathcal{O}\left(K \left(2^DJ +1\right) N_R N_T+K N_R\right)$, which linearly increases as the number of RIS elements increases.

\subsection{Asymptotic Results}
{In this part, we give an asymptotic analysis framework to rigorously quantify the achievable gain of the proposed RIS design compared with traditional passive RIS in the absence of external power feed.}

{We denote by $P_{\text{old}}$ and $P_{\text{new}}$ the reflected power of the traditional passive RIS and the proposed self-amplifying RIS, respectively.} The effective power gain provided by the proposed RIS, denoted by $P_{\text{gain}}$, is defined by the ratio as $P_{\text{new}}/P_{\text{old}}$. To derive palpable conclusions, we make the following assumptions for simplicity. First, the number of BS antenna is set to 1 and the direct link is omitted to fully embody the power gain of RIS. {Second, the uniform spherical wave (USW) region is considered with the same amplitude but non-linear phase distribution\cite{USW}, as opposed to non-uniform spherical wave (NUSW) with different gains among array elements. The critical point to differentiate USW and NUSW is defined as the uniform-power distance (UPD), which can be expressed by $\sqrt{\Gamma_{\rm th}^{\frac{2}{3}}/(1-\Gamma_{\rm th}^{\frac{2}{3}})}\frac{L_R}{2}$, where $\Gamma_{\rm th}$ and $L_R$ are minimum threshold for the power ratio and the diagonal dimension of RIS, respectively\cite{USW_define}.} Under the above conditions, we denote by $h$ the uniform channel gain between BS and RIS elements. According to Section IV-B, the proportion of reflective elements, denoted by $p$, can be expressed by $p={\chi2^D}/{\pi}$ with $p \in [0,1]$. Also, the harvested energy from uplink is negligible compared with that of downlink, so we omit it in power budget.

{Under such circumstance, based on the power consumption relation in Eq.~\eqref{power_con}, we can derive the following relation:
\begin{equation}
    \begin{aligned}
    \underbrace{\left((1-p) \varrho^{\rm a} \eta_1 \eta_2 N_R h\right) ^2}_{\text{Harvested power}}&=\underbrace{N_R p \left(\rho^2-1\right) h^2}_{\text{Amplifying consumption}}\\
    &~~~~~~+\underbrace{P_c +P_{DC}+f(D)N_R}_{\text{Circuit consumption}}, \label{relation}        
    \end{aligned}
\end{equation}
where the left and right parts of Eq.~\eqref{relation} denote the harvested power by absorptive elements and consumed power including amplifying circuits, RIS controller, and element consumption, respectively, while meeting 
\begin{equation}
   {\left((1-p) \varrho^{\rm a} \eta_1 \eta_2 N_R h\right) ^2}>P_c +P_{DC}+f(D)N_R \label{power-budget}
\end{equation}
to achieve the capability of zero external power. Then, the value of $\rho^2$, determined by the available power for amplification circuits, can be derived as follows:
\begin{equation}
    \rho^2=\frac{\left(\varrho^{\rm a} \eta_1 \eta_2(1-p)\right) ^2 N_R }{ p}+\frac{P_c +P_{DC}+f(D)N_R}{N_R p h^2}, \label{rho}
\end{equation}
For traditional passive RIS, the power gain
$P_{\text{old}}$ has the relation as $P_{\text{old}}\propto \left(N_Rh\right) ^2$ with respect to RIS elements. Similarly, the power gain for the proposed RIS $P_{\text{new}}$ has the relation as follows:
\begin{equation}
    P_{\text{new}}\propto\left(\rho N_R ph\right) ^2.
\end{equation}
Then, $P_{\text{gain}}$ can be expressed as follows:
\begin{equation}
    P_{\text{gain}}= p ^2 \rho^2. \label{gain}
\end{equation}}
By substituting Eq.~\eqref{rho} to Eq.~\eqref{gain}, $P_{\text{gain}}$ can be rewritten as follows:
\begin{equation}
    P_{\text{gain}}=p \left(\varrho^{\rm a} \eta_1 \eta_2(1-p)\right) ^2 N_R + \frac{\left(P_c +P_{DC}+f(D)N_R\right)p }{N_R h^2}. \label{Powergain}
\end{equation}
Hence, to find the optimal value of $p$ for maximizing $P_{\text{gain}}$, let the derivative of $P_{\text{gain}}$ corresponding to $p$ equal to zero as follows:
{\begin{equation}
    \begin{aligned}
    \frac{d P_{\text{gain}}}{d p}&=\left(\varrho^{\rm a} \eta_1 \eta_2\right)^2  N_R \times\\ &\left(3 p^2-4p+1+\frac{P_c +P_{DC}+f(D)N_R}{ \left(N_R h \varrho^{\rm a} \eta_1 \eta_2\right)^2}\right)=0.        
    \end{aligned}
\end{equation}}
Then, we can have
\begin{equation}
    p=\frac{2}{3}\pm\frac{\sqrt{N_R^2 h^2-P_c -P_{DC}-f(D)N_R}}{3N_R h \varrho^{\rm a} \eta_1 \eta_2}.
\end{equation}
{By verifying the second derivative, the optimal value of $p$, denoted by $p^{\text{opt}}$, can be expressed as follows:
\begin{equation}
    \begin{aligned}
       p^{\text{opt}}=\frac{2}{3}-\frac{\sqrt{N_R^2 h^2-P_c -P_{DC}-f(D)N_R}}{3N_R h \varrho^{\rm a} \eta_1 \eta_2}.
       \label{optimal_p}
    \end{aligned}
\end{equation}
To ensure the sustainability, by substituting Eq.~\eqref{power-budget} to Eq.~\eqref{optimal_p}, the following constraint should be satisfied:
\begin{equation}
    \begin{aligned}
    p^{\text{opt}}&<\frac{2}{3}-\\
    &\frac{\sqrt{\left(1-\left(\varrho^{\rm a} \eta_1 \eta_2\left(1-p^{\text{opt}}\right) \right)^2\right) \left(P_c +P_{DC}+f(D)N_R\right)  }}{3N_R h \left(\varrho^{\rm a} \eta_1 \eta_2\right)^2\left(1-p^{\text{opt}}\right) }.    \label{power-constraint}
    \end{aligned}
\end{equation}
When $N_R$ is large, $p^{\text{opt}}$ in Eq.~\eqref{optimal_p} can be further reduced to
\begin{equation}
    p^{\text{opt}}=\frac{2}{3}-\frac{1}{3 \varrho^{\rm a} \eta_1 \eta_2}.
\end{equation}}
{Hence, the optimal value of $\chi$, denoted by $\chi^{\text{opt}}$, which determines the proportion of the pre-defined $\mathcal{D}$ and $\mathcal{I}$ regions, can be expressed as follows:
\begin{equation}
   \chi^{\text{opt}}= \frac{\left(2 \varrho^{\rm a} \eta_1 \eta_2-1\right) \pi}{3 \varrho^{\rm a} \eta_1 \eta_2 2^D}.
\end{equation}
Additionally, under such circumstance, the maximum asymptotic achievable gain of the proposed RIS, denoted by $P_{\text{gain}}^{\text{max}}$, can be expressed as follows:
\begin{equation}
    P_{\text{gain}}^{\text{max}}= \left(\frac{2}{3}\rho-\frac{1}{3 \varrho^{\rm a} \eta_1 \eta_2} \rho\right) ^2. 
\end{equation}}
According to the aforementioned analysis, the following \emph{Remark 3} can be summarized:
{\begin{remark}
    When $N_R$ is large, the optimal proportion of $\mathcal{D}$ and $\mathcal{I}$ regions approaches $\chi^{\text{opt}}=\frac{2}{3}-\frac{1}{3 \varrho^{\rm a} \eta_1 \eta_2}$, where the maximum achievable gain can be derived subsequently as $\left(\frac{2}{3}\rho-\frac{1}{3 \varrho^{\rm a} \eta_1 \eta_2} \rho\right) ^2$. This indicates that in the large $N_R$ region, the conversion efficiency of the EH module and the absorption efficiency of the absorptive elements will jointly determine the optimal proportion of the two regions. 
\end{remark}}

\vspace{-0.3 cm}

\section{Numerical Results}
In this section, we evaluate the performance of the proposed scheme. The simulation parameters are set as follows unless specified otherwise. The operation frequency is set to 30 GHz, the center of Tx is located at $\left(0, 0, 15\right)$ equipped with two antennas spacing $\lambda/2$, the UE is located at $\left(5, 5, 1.5\right)$, the RIS consists of $50\times 50$ elements with a spacing of $\lambda/2$, which is deployed in $x-y$ plane and its center is placed at the origin of Cartesian coordinate system. The transmit power and noise power are set to 30 dBm and -90 dBm, respectively, the absorptive efficiency as well as the charge and discharge efficiency are set to $\varrho ^{\rm a}=\eta_1 =\eta_2 =0.9$, the maximum reflective coefficient is set to $\rho=10$ to ensure active RISs operate in linear amplification region, the antenna gains of Tx and UE are set to 15 dBi and 0 dBi, respectively. {Under the above parameters, by setting $\Gamma_{\rm th}=0.95$, the region satisfying uniform spherical wave condition is $[0.94,25]$ m, which is suitable for verifying the asymptotic results.} In addition, to verify the superiority of the proposed element-wise strategy, the PS\cite{Selfris4}, TS\cite{Selfris3}, and ES \cite{Selfris1} strategies are selected as benchmark schemes.

\begin{figure}[t]
    \centerline{\includegraphics[width=9cm]{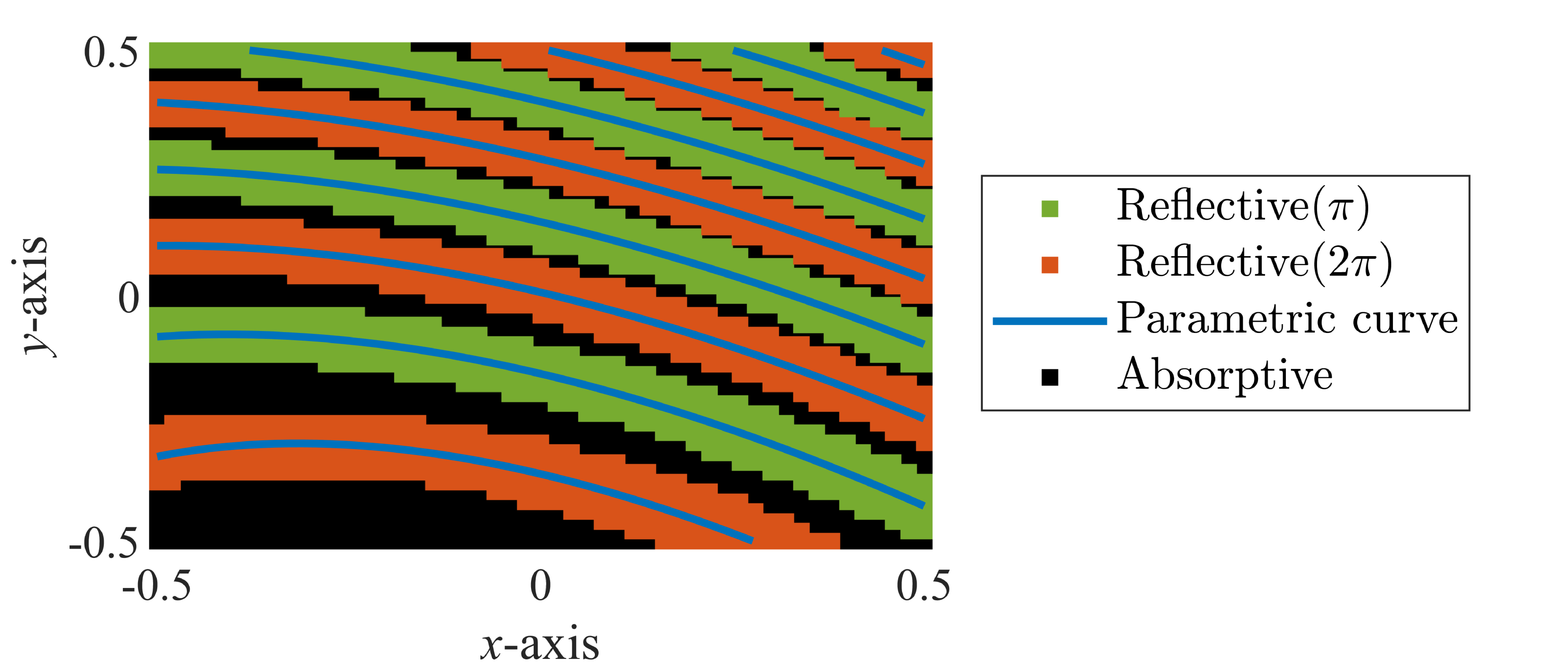}}
        \caption{{One example of joint reflective and absorptive coefficients design with 1-bit phase resolution.}}
        \label{1-bit-RIS}
        \vspace{-0.5 cm}
\end{figure}

\begin{figure}[t]
    \centerline{\includegraphics[width=9cm]{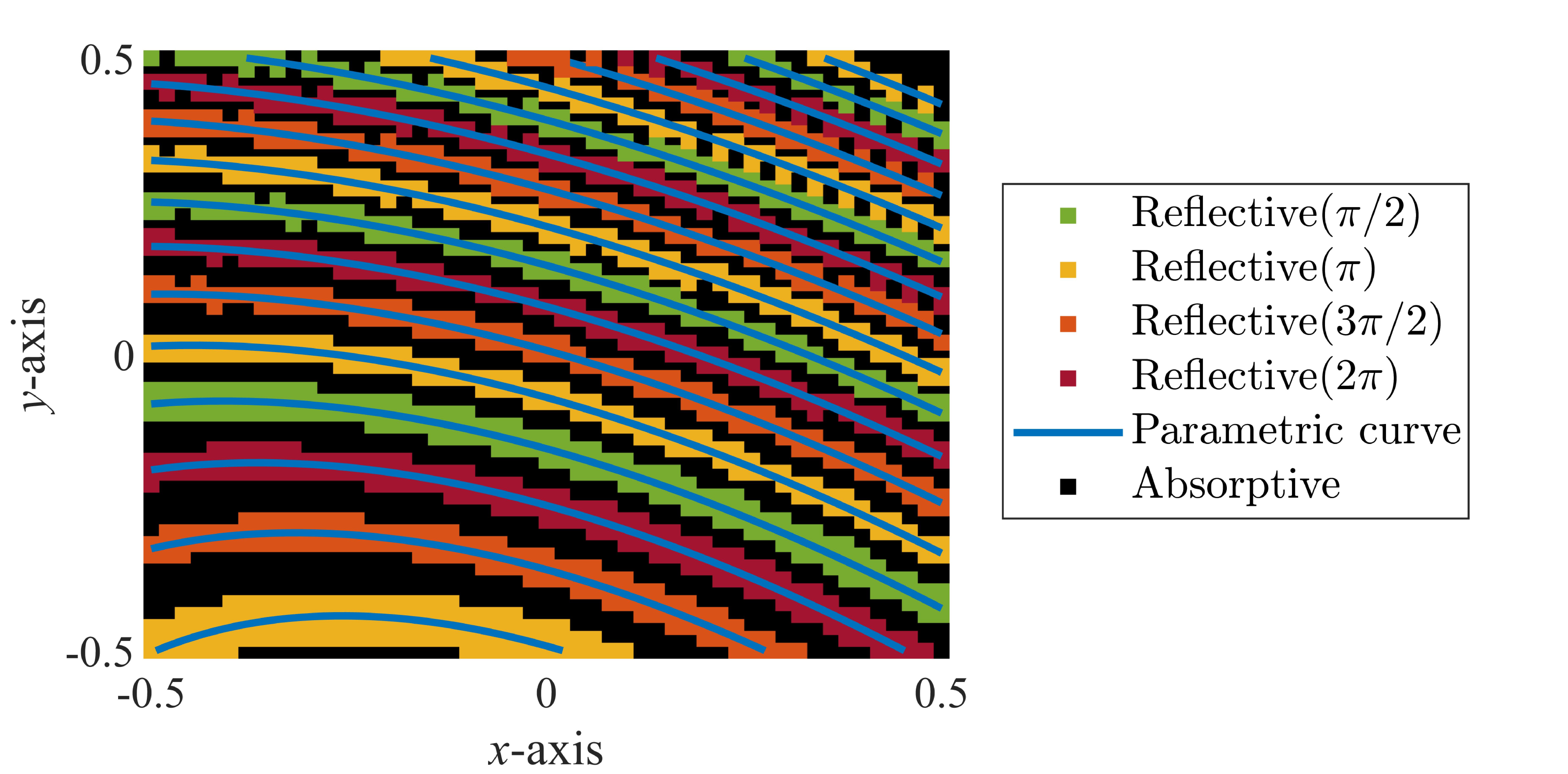}}
        \caption{{One example of joint reflective and absorptive coefficients design with 2-bit phase resolution.}}
        \label{2-bit-RIS}
        \vspace{-0.5 cm}
\end{figure}
Figure \ref{1-bit-RIS} illustrates one example of joint reflective and absorptive coefficients design with 1-bit phase resolution. Under 1-bit phase resolution, the parametric lines represent the phase distributions of incident waves for the traditional Fresnel zone case. We denote by $\pi$ and $2\pi$ two states of reflective elements, respectively. As shown in Fig.~\ref{1-bit-RIS}, the parametric lines form parts of uninterrupted ellipses with obvious curvature in the near-field region. According to the element selection criterion, the elements near $\mathcal{I}$ and $\mathcal{D}$ regions are configured as absorptive and reflective elements, respectively. Hence, the shapes formed by RIS elements also behave similarly to ellipses. Also, it can be observed that the number of absorptive elements gradually decreases as the index of Fresnel zone ellipsoids increases. This is because the Fresnel radius decreases as the index of Fresnel zone ellipsoids increases. Figure \ref{2-bit-RIS} illustrates one example of joint reflective and absorptive coefficients design with 2-bit phase resolution, where the phase shift is chosen from the set $\left\{\pi/2, \pi, 3\pi/2, 2\pi\right\} $. Under 2-bit phase resolution, the parametric lines become denser as the proposed Fractional Fresnel zone model divides the traditional Fresnel zone into more pieces to match the phase resolution of RIS. Based on Eq.~\eqref{reflective}, the phase shifts of reflective elements are configured according to the index of the fractional Fresnel zone.

\begin{figure}[t]
    \centerline{\includegraphics[width=8cm]{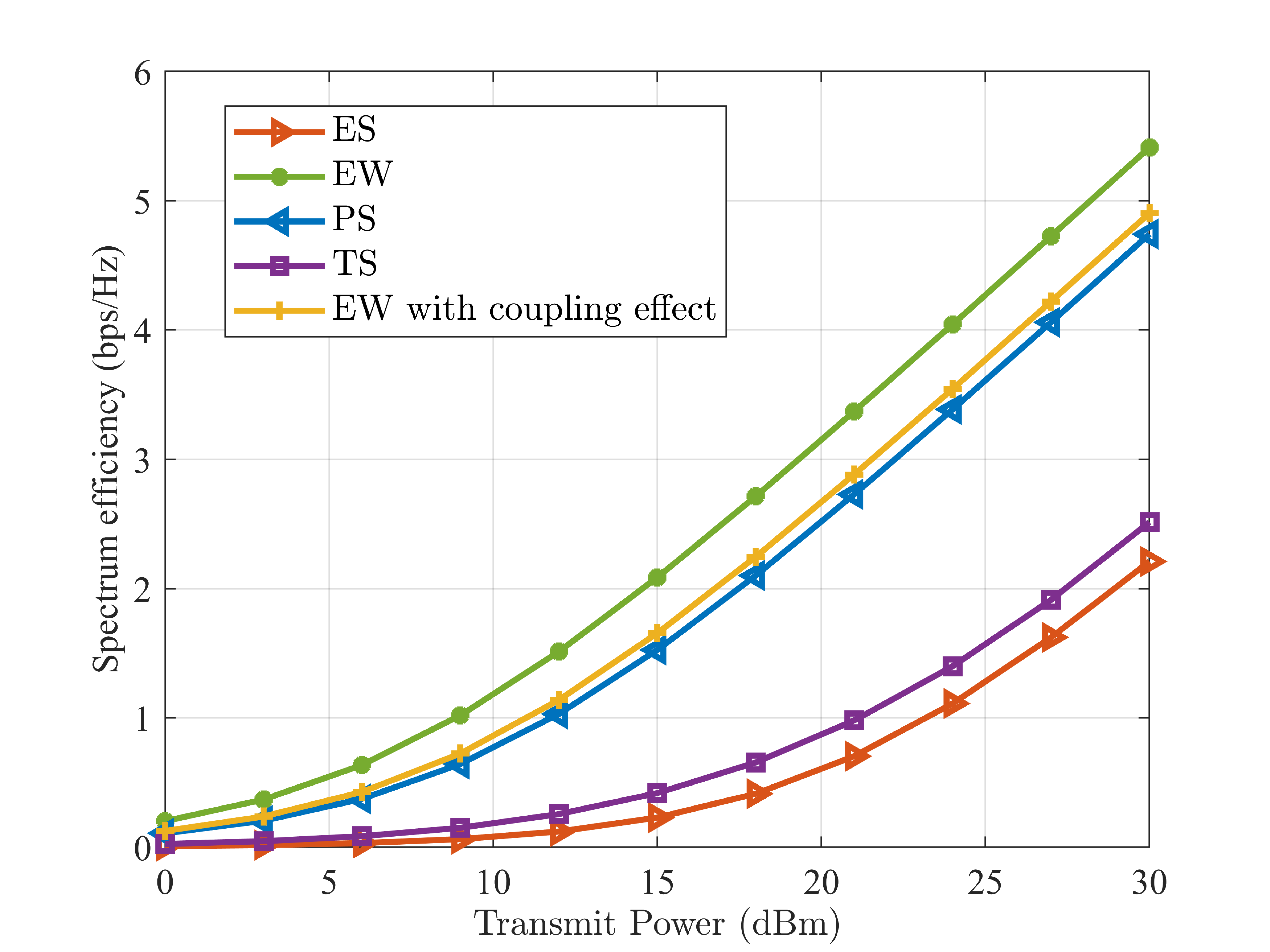}}
        \caption{{The spectrum efficiency versus the transmit power in dBm with 1-bit phase shift resolution.}}
        \label{capacity_viaP}
        \vspace{-0.5 cm}
\end{figure}
{Figure \ref{capacity_viaP} depicts the spectrum efficiency versus the transmit power with 1-bit phase shift resolution under different strategies, including the element-wise, PS, TS, and ES strategies, and also investigate the performance loss caused by coupling effect. In each condition, we determine the optimal PS factor, time splitting factor, and the proportion of elements allocated for EH for each transmit power to make a fair comparison. As shown in Fig.~\ref{capacity_viaP}, the performance of the proposed EW strategy slightly surpasses that of the PS strategy, and greatly surpasses that of the TS, and ES strategies. Generally, with perfect phase shift case, the PS strategy is the most effective strategy to harness the reflected gain induced by active elements. However, considering the discrete phase resolution, its performance will deteriorate as it mixes elements from different regions. Hence, the proposed EW strategy compensates the phase deviations by element selection and leverages the elements in $\mathcal{D}$ region with large effective gain for reflecting, which captures the element selection gain to achieve better performance in the low $D$ region. Additionally, in the presence of coupling effect, the performance of the proposed EW strategy decreases. This is because the coupled phase and amplitude responses affect the effective gain of different regions, leading to a performance degradation.
}

\begin{figure}[t]
    \centerline{\includegraphics[width=8cm]{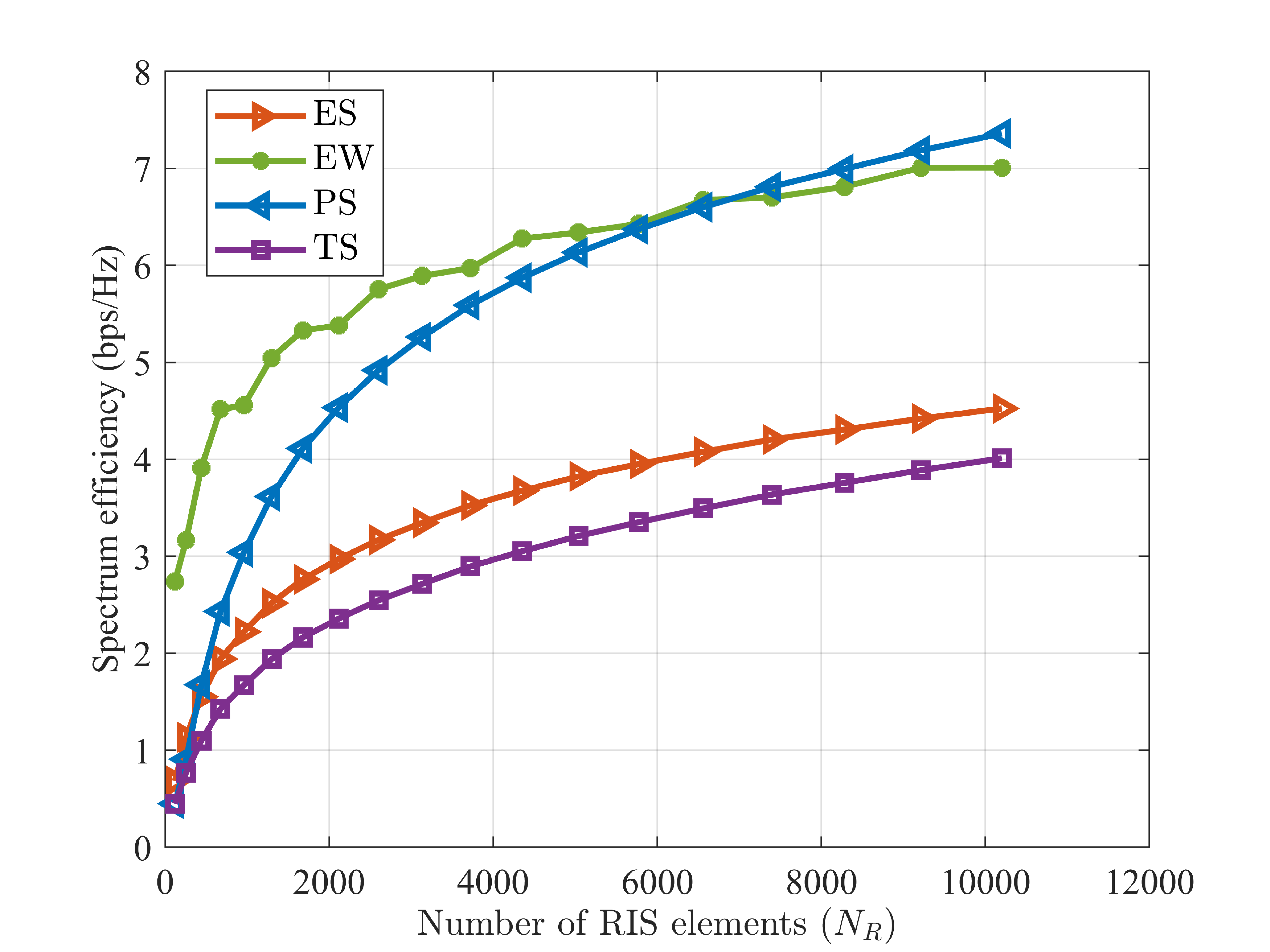}}
        \caption{The spectrum efficiency versus the number of RIS elements $N_R$.}
        \label{capacity_viaMN}
        \vspace{-0.5 cm}
\end{figure}
Figure \ref{capacity_viaMN} presents the spectrum efficiencies of different strategies versus the number of RIS elements $N_R$. As shown in Fig.~\ref{capacity_viaMN}, the spectrum efficiencies of all strategies increase as the number of RIS elements increases. The performances of TS and ES strategies still maintain a low level compared with ES and EW strategies, which is consistent with Fig.~\ref{capacity_viaP}. {Additionally, in the low $N_R$ region, the gap between the reflected gains of proposed EW strategy and PS strategy is large and gradually decreases as $N_R$ increases. Then, with the increase of RIS elements, the spectrum efficiency of PS strategy increases rapidly and finally surpasses that of the EW strategy. The reason is that the PS strategy exploits all the RIS elements to simultaneously harvest and reflect the signals. When the number of RIS element is very large, the power gain induced by enormous RIS elements will surpass the gain of element selection.} 

\begin{figure}[t]
    \centerline{\includegraphics[width=8cm]{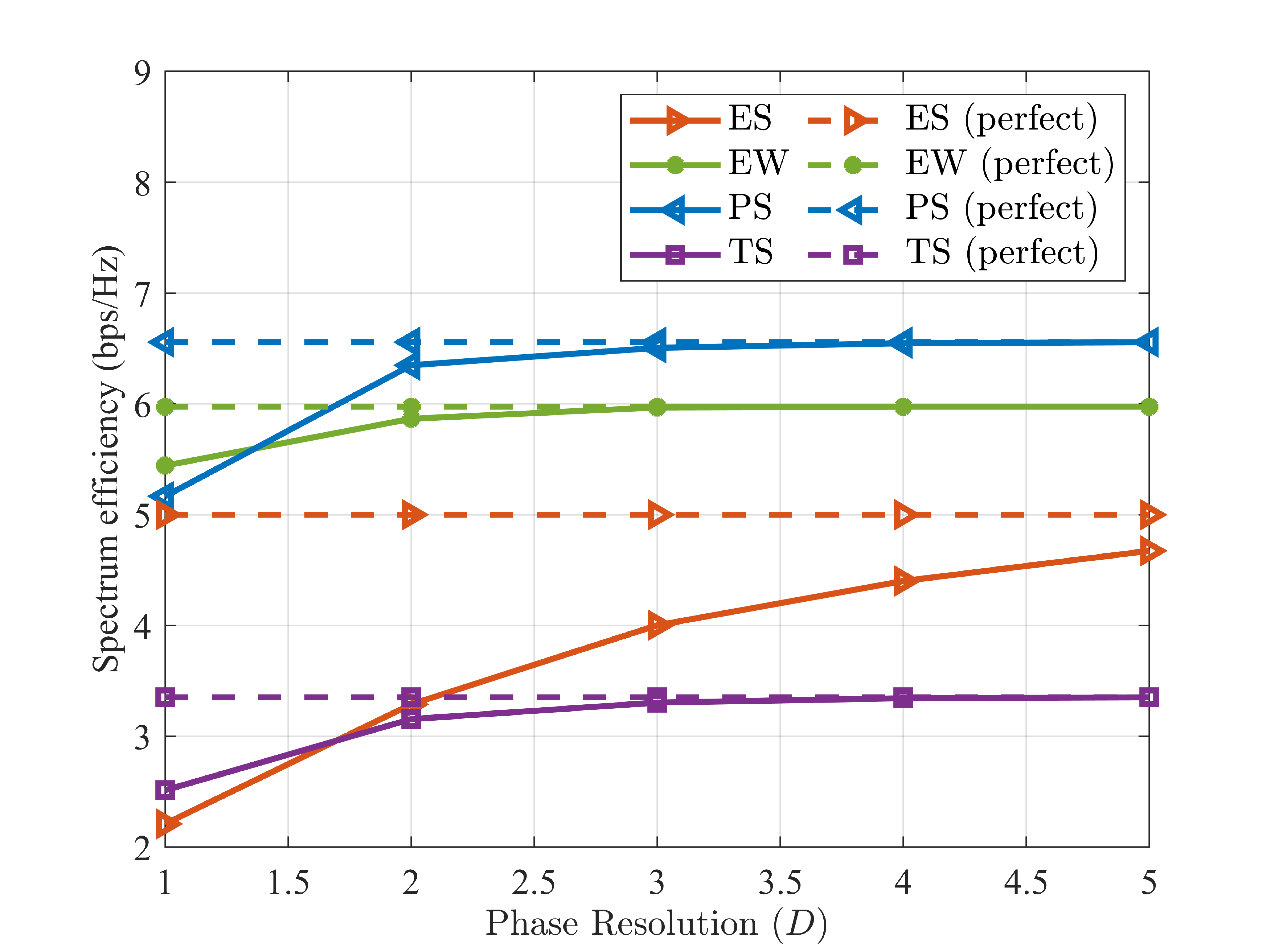}}
        \caption{{The spectrum efficiency versus the phase resolution $D$.}}
        \label{capacity_viaD}
        \vspace{-0.5 cm}
\end{figure}
Figure \ref{capacity_viaD} shows the spectrum efficiencies of different strategies versus the phase resolution $D$ and the corresponding perfect phase shift case. From Fig.~\ref{capacity_viaD}, it can be seen that the performances of all strategies increase as the phase resolution increases. Apart from the ES strategy, the spectrum efficiencies corresponding to other strategies basically approach the perfect phase shift case when $D=3$, indicating that the low phase shift resolution is sufficient to provide effective reflected gain. However, limited by the divided time for transmission in TS strategy, the performance of ES strategy gradually surpasses that of the TS case as $D$ increases, where the ES strategy has a large margin for capacity improvement, indicating the potential of element selection design. {Additionally, as the phase resolution $D$ increases, the PS strategy outperforms the EW strategy due to the power gain induced by $N_R$ elements with simultaneously harvesting and reflecting capabilities. Nevertheless, such desired gain necessitates complicated hardware overhead to achieve. In PS strategy, to enable both EH and reflecting capabilities, each element should be equipped with one power divider and two phase shifters, which totally needs $N_R$ power dividers and $2N_R$ phase shifters\cite{Autonomous_RIS2}. By contrast, the proposed element-wise design only needs $N_R$ switches and one phase shifter with acceptable power gain loss when $D$ is larger than 1. Therefore, the proposed EW scheme can greatly reduce the interconnect overhead. More importantly, when $D=1$, the element-wise design can achieve better performance than other strategies, providing a competitive solution for specific scenarios, such as the hardware-limited case. Moreover, in the proposed distributed estimation design, the allocation of power splitting coefficients in PS strategy and UE's location will jointly affect the result of the power indicator, leading to more complicated control to sense UE's location.}

%Additionally, as the phase resolution $D$ increases, the PS strategy outperforms the EW strategy due to its capability of simultaneously harvesting and reflecting. Nevertheless, the ideal function of PS strategy is hard to achieve, which necessitates complicated hardware design. Hence, the proposed EW RIS is an effective solution with low hardware complexity.

\begin{figure}[t]
    \centerline{\includegraphics[width=8cm]{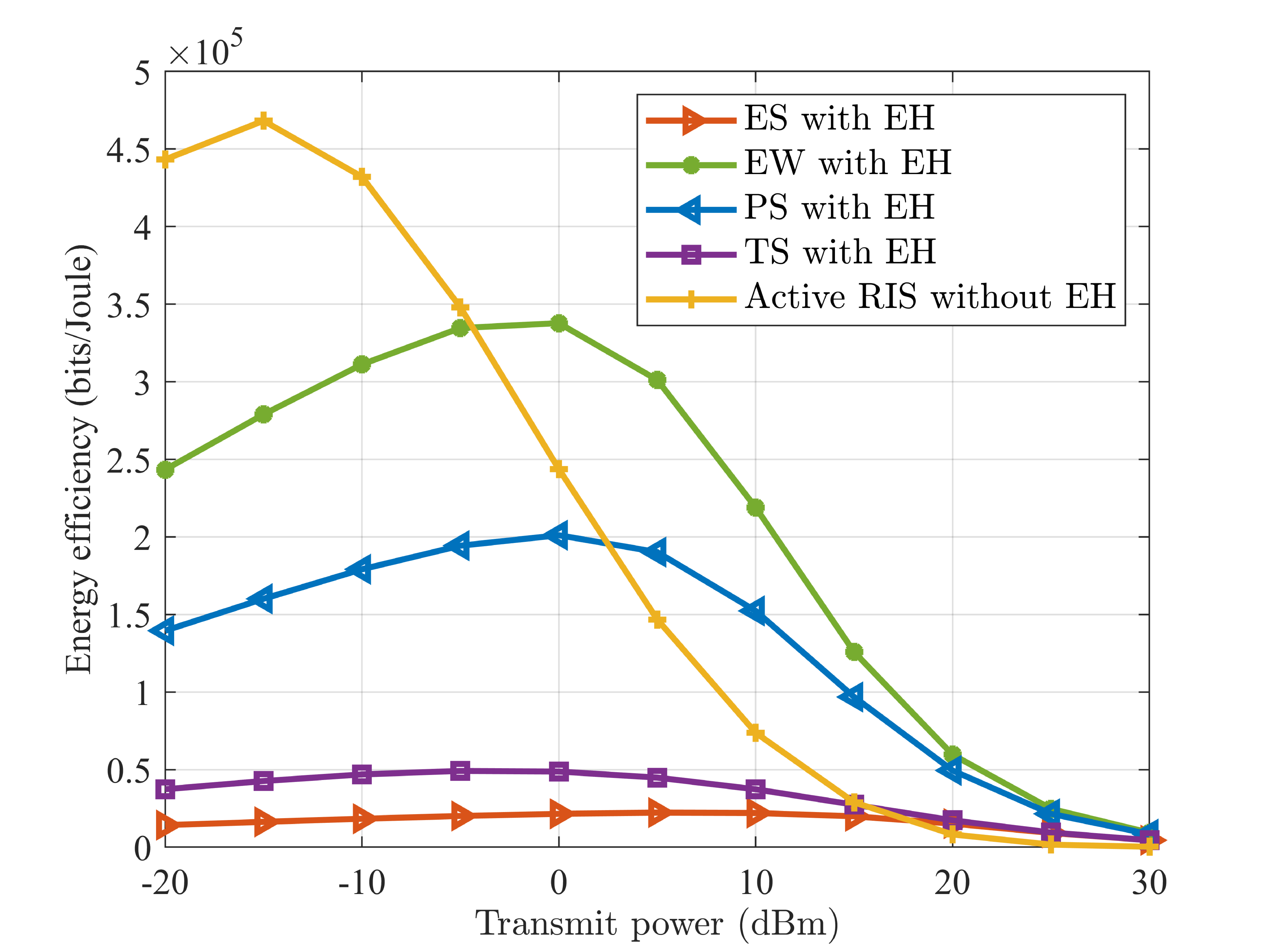}}
        \caption{{The energy efficiency versus the transmit power in dBm under 1-bit resolution.}}
        \label{EE_via_Pt}
        \vspace{-0.5 cm}
\end{figure}
{Figure \ref{EE_via_Pt} illustrates the energy efficiencies (EEs) of different sustainable schemes with EH and the benchmark scheme without EH under $\rho=10$. Therein, the benchmark active RIS is fully powered by an external and stable energy supply, where the power allocated to the BS and RIS is equal. As can be seen from Fig.~\ref{EE_via_Pt}, EEs of all schemes firstly increase and subsequently decrease as transmit power increases. Also, the peak of the benchmark without EH arrives earlier than that of the schemes with EH, exhibiting better performance. This is because in the low transmit power region, the harvested power for these schemes with EH is limited, which restricts the superiority of active RIS. Nevertheless, as transmit power increases, performance gap narrows and finally EEs of the schemes with EH surpass that of the benchmark scheme, which stems from the limitation of $\rho$. When the power at RIS in the benchmark scheme is large, the reflected gain is mainly limited by the nonlinear effect of amplifier. Additionally, the proposed EW strategy exhibits optimal EE over other sustainable schemes. The performance increment is achieved by leveraging element selection to remedy the quantization error.}

\begin{figure}[t]
    \centerline{\includegraphics[width=8cm]{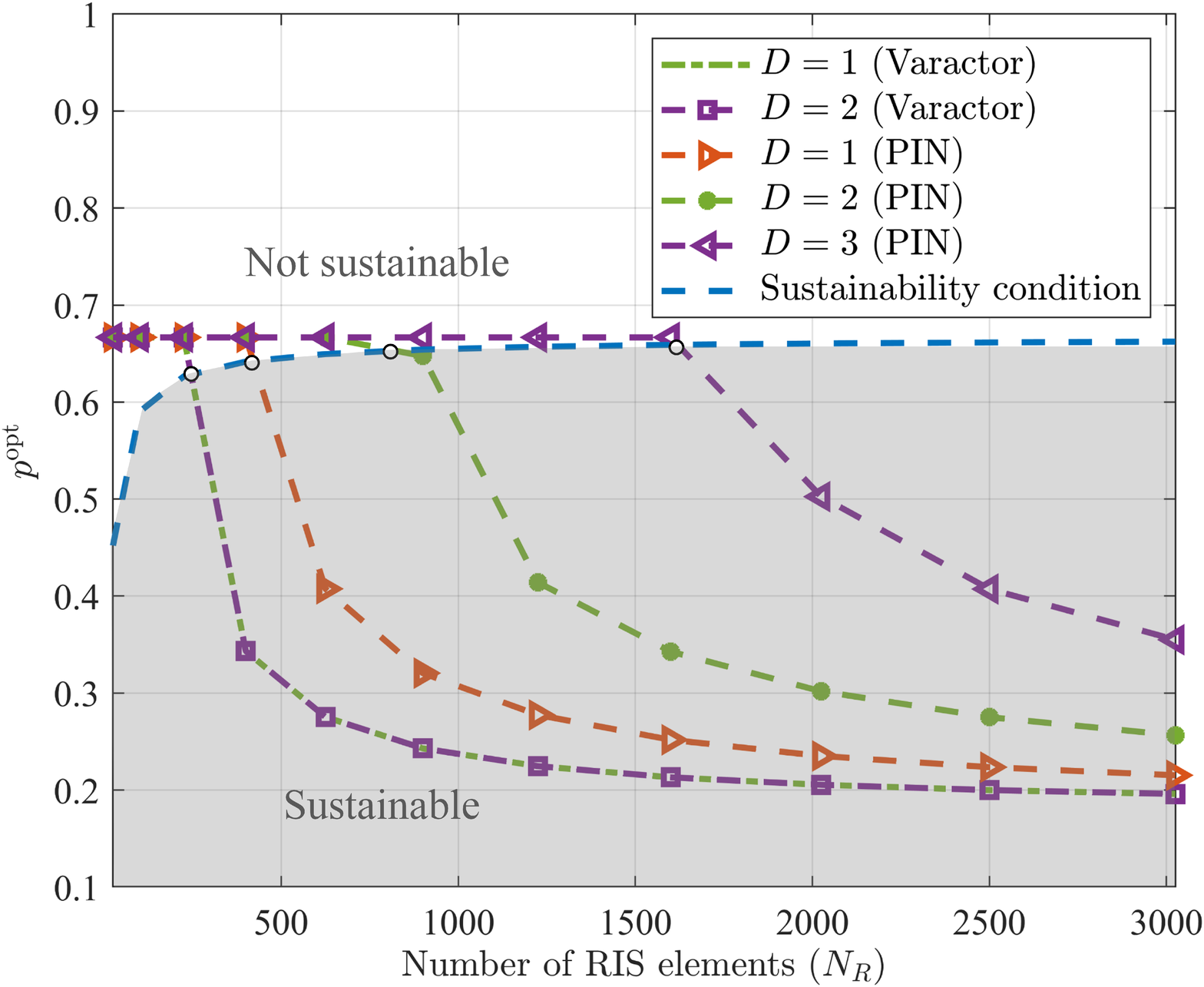}}
        \caption{{The optimal proportion of RIS element versus the number of RIS elements with different $D$.}}
        \label{p_opt}
        \vspace{-0.5 cm}
\end{figure}
Figure \ref{p_opt} plots the optimal proportion of reflective RIS elements versus the number of RIS elements with different phase resolutions $D$. The power consumption satisfies $f(D) \propto D$ for PIN diode-based elements while remaining a constant for varactor-based elements. As shown in Fig.~\ref{p_opt}, the $p^{\text{opt}}$ of all cases exists a stable region when $N_R$ is small and gradually approaches a same value when $N_R$ is large. {Therein, in the small $N_R$ region, the harvested power by absorptive elements is not sufficient to supply the power consumption of systems in Eq.~\eqref{power-budget}, which is impacted by the power budget of the circuits determined by $D$ and element type, leading to a invalid value of $p^{\text{opt}}$. It is noteworthy that the constraint in Eq.~\eqref{power-constraint} generate a sustainability condition, where the region upper the sustainability condition cannot satisfy the sustainable function. Hence, the intersection between the $p^{\text{opt}}$ and sustainability condition is the real boundary that determines the sustainability. Then, as $N_R$ increases, the sustainability can be satisfied when $p^{\text{opt}}$ is less than the intersection point. Finally, in the large $N_R$ region, $p$ of all cases is dominated by $N_R$ and the sustainability condition can be easily satisfied so that $p^{\text{opt}}$ is determined by Eq.~\eqref{optimal_p}.} Also, as the phase resolution $D$ increases, $p^{\text{opt}}$ gradually decreases to capture the maximum effective gain. As the power consumption of PIN diodes typically surpasses that of varactor, the corresponding $p^{\text{opt}}$ of PIN is larger than that of varactor under the same phase resolution. 

\begin{figure}[t]
    \centerline{\includegraphics[width=8cm]{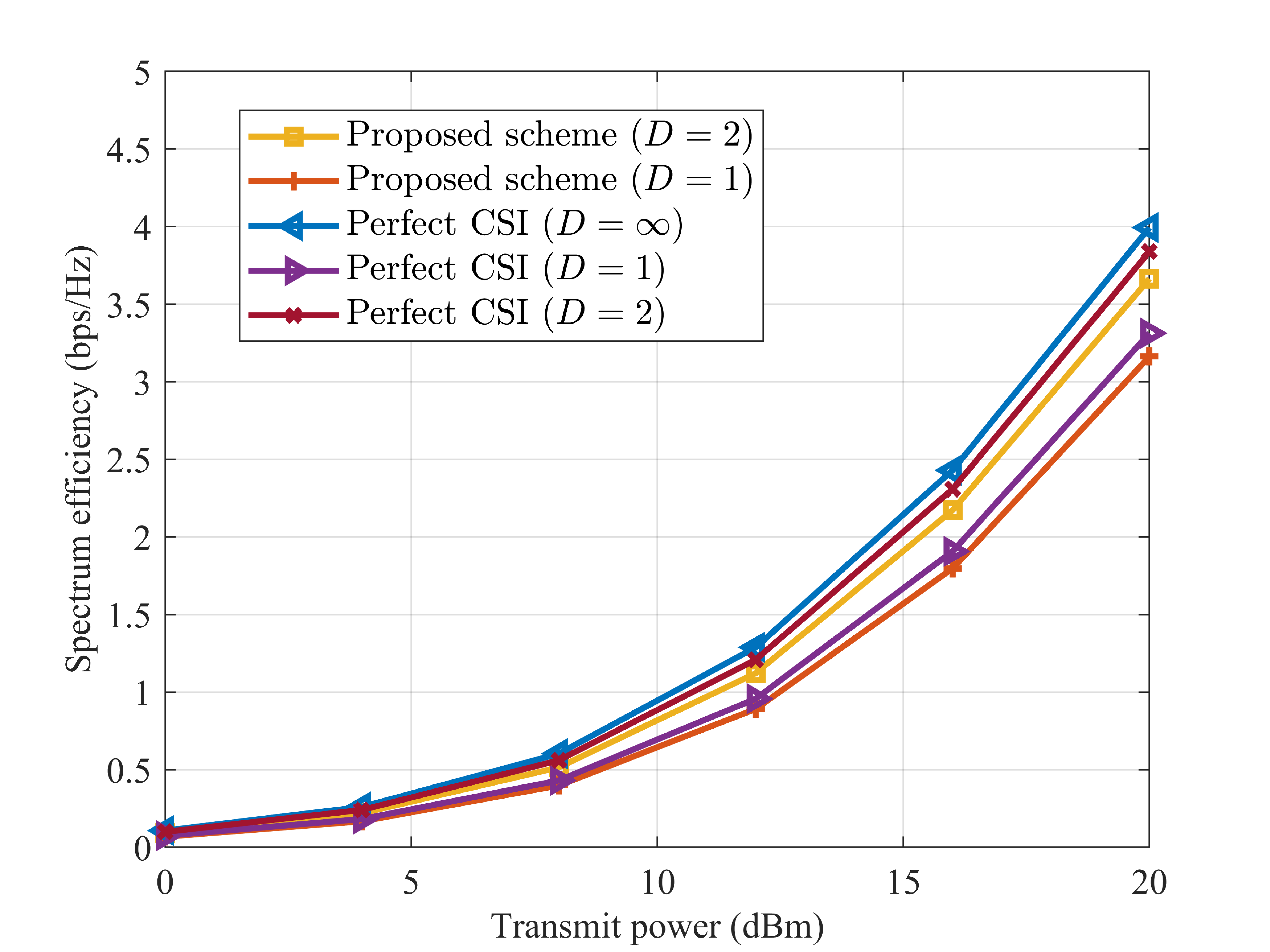}}
        \caption{{The spectrum efficiencies of CSI-driven scheme and the proposed scheme versus the transmit power in dBm.}}
        \label{CSI}
\vspace{-0.5 cm}        
\end{figure}

{{Figure \ref{CSI} depicts the spectrum efficiencies of CSI-driven scheme with the alternating optimization algorithm and the proposed scheme under different $D$.} As shown in Fig.~\ref{CSI}, under the same phase resolution, the CSI-driven scheme with perfect CSI achieves better performance than the proposed location-based scheme due to its joint design of beamforming at BS and RIS. Nevertheless, the performance increment of CSI-driven scheme is minor, which is unworthy of investing complicated estimation complexity and large overhead\cite{Xiaozheng_mag}. From the physical perspective, the joint design of beamforming at BS and RIS induces beam energy allocation towards BS and RIS. Although the active RIS can remedy the double fading effect, it only shows the great promotion in the NLoS scenario\cite{Active-RIS-measurement}. It is noteworthy that the path gain of active RIS is about 11.4 dB less than that of the direct link under the above setting. Hence, as the strength of LoS link is typically greater than that of the reflected link from RIS, the optimal beam allocation of BS precoding leads to a stronger beam towards UE and a weaker beam towards RIS. In the proposed distributed beamforming scheme, the MRT precoding directly align the beam towards the UE, which leads to a sub-optimal beam allocation result that is close to the optimal one. This is why the gap between optimal joint beamforming and the proposed scheme is insignificant. Also, the spectrum efficiency of the proposed scheme with $D=2$ nearly approaches the perfect CSI case with perfect phase resolution, indicating that exploiting RIS with low phase resolution is sufficient to balance the hardware overhead and performance.}

\begin{figure}[t]
    \centerline{\includegraphics[width=8cm]{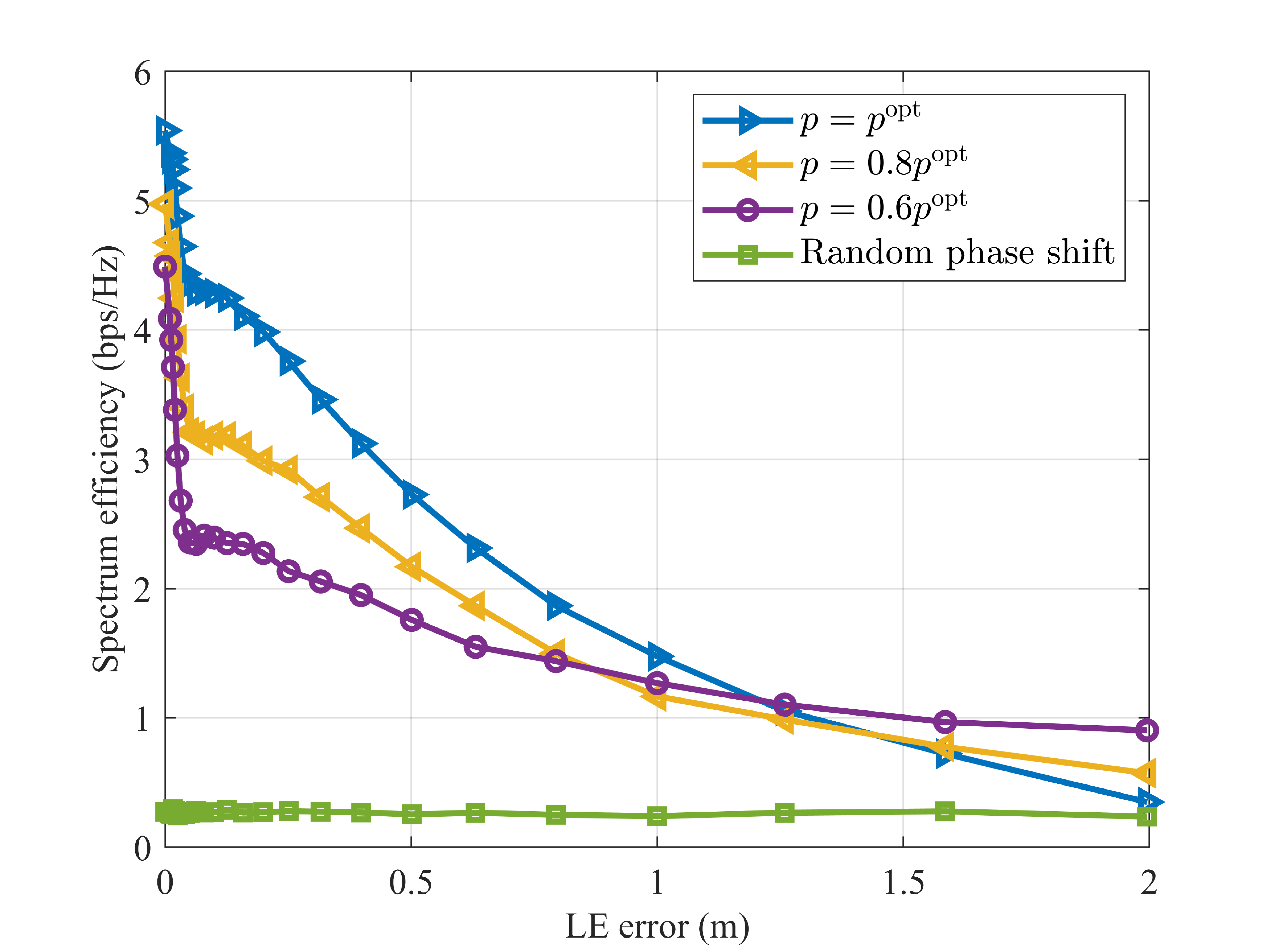}}
        \caption{{The spectrum efficiency of the proposed scheme versus LE error.}}
        \label{LEerror}
 \vspace{-0.5 cm}       
\end{figure}

\begin{figure}[t]
    \centerline{\includegraphics[width=8cm]{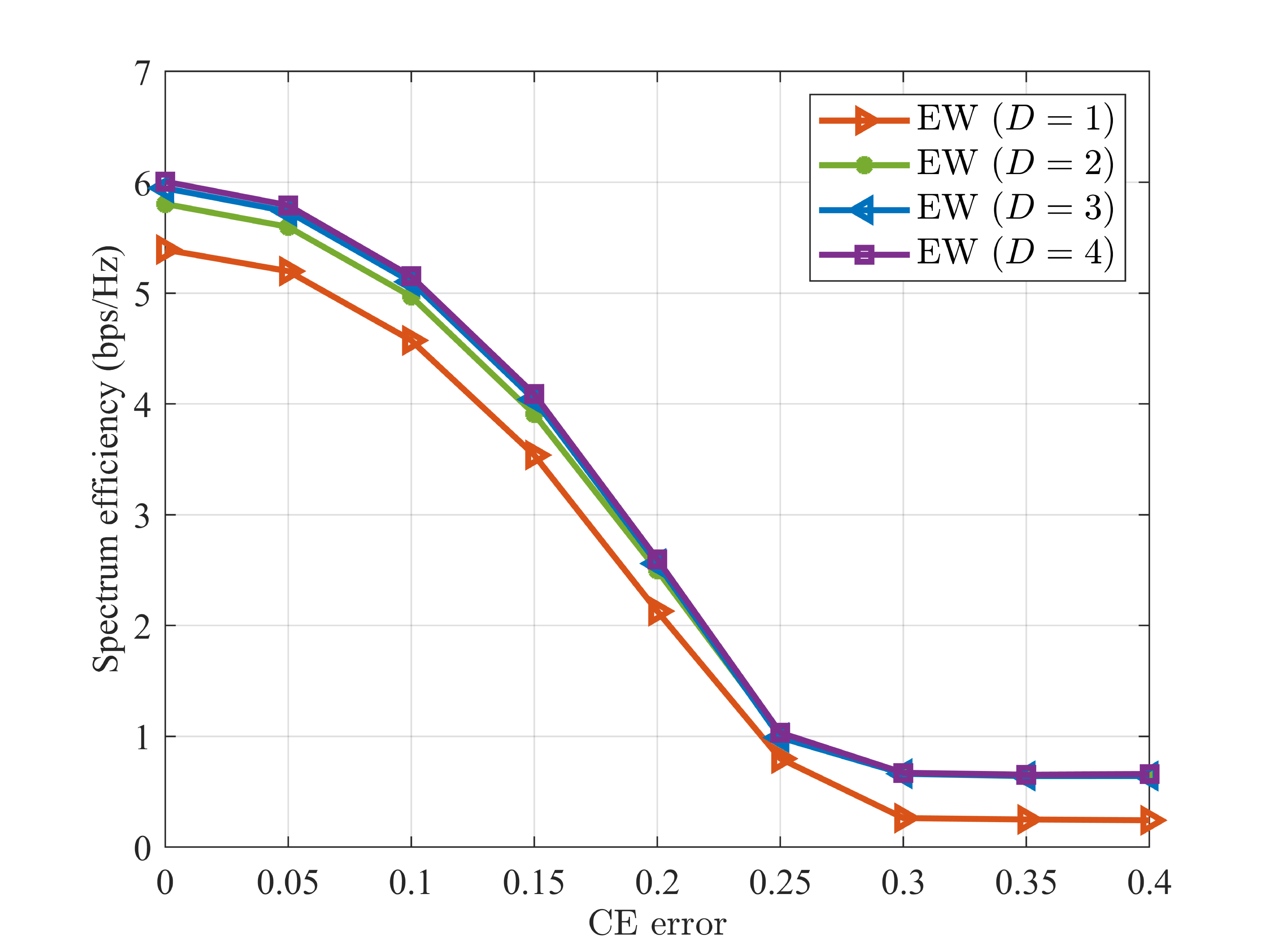}}
        \caption{{The spectrum efficiency of the proposed scheme versus CE error.}}
        \label{CEerror}
\vspace{-0.5 cm}        
\end{figure}

{Figures \ref{LEerror} and \ref{CEerror} show the spectrum efficiencies of the proposed scheme versus LE error with different proportions of reflective elements $p$ and statistical CE error with different $D$. It can be observed from Fig.~\ref{LEerror} that as location error increases, the performance of proposed scheme first rapidly degrades in the low error regime, and later slowly degrades until approaching the random phase shift case, which contains a relatively stable interval. When location error is small, the performance with higher $p$ surpasses that of with lower value and achieves optimal performance with $p=p^{\rm opt}$. The small $p$ leads to fewer reflective elements, where the performance is limited by the maximum amplifying factor $\rho$. Nevertheless, as $p$ decreases, the rate of decline slows down and the stable interval extends. {This phenomenon is consistent with the previous analysis, wherein the region exhibit steadfast constructive interferences compared with those within $\mathcal{I}$ region, such elements are less influenced by LE error.} Hence, the robustness of the proposed scheme can be enhanced by sacrificing part of the reflected gain. Additionally, it can be seen from Fig.~\ref{CEerror} that as the CE error increases, the spectrum efficiencies of the proposed scheme first slowly degrade and then rapidly degrade until approaching a stable value. This trend stems from the configuration criteria in Eq.~(38). The increase of CE error leads to the increase of phase deviation between two transmit antennas. When CE error is small, the phase difference value between the two transmit antennas is small so that the strength of the composed signal is strong. As CE error increases, the phases of two signals will be in destructive interference, resulting in interference cancellation with low strength. In addition, the spectrum efficiency with higher $D$ exhibits better performance while the performance gap becomes smaller as $D$ increases, which is consistent with Fig.~\ref{capacity_viaD}.
}

\vspace{-0.5 cm}
\section{Conclusions}
In this paper, we designed a new RIS structure and proposed a distributed location-aided beamforming scheme considering the discrete phase shift constraint for CSI-limited RIS-assisted near-field communications. Specifically, to tackle the new problem arising from discrete phase shift, we designed the element-wise RIS with low hardware resources. Then, based on Fresnel diffraction theory, we constructed the mapping from locations in space-domain to phase distributions of waves in phase-domain. Then, by considering the phase deviations induced by discrete phase shift, we revealed the priority of elements for harvesting and reflecting, with which the distributed beamforming design with the phase of determine-then-align was proposed. The asymptotic analysis showed that the optimal gain can be achieved with a fixed proportion of reflective elements when RIS is large. Finally, simulations indicated that under the discrete phase shift constraint, the performance of the proposed EW strategy and corresponding beamforming can effectively outperform other existing protocols, verifying its effectiveness in practical scenarios.

\vspace{-0.2 cm}
\bibliography{References}
\bibliographystyle{ieeetr}

\end{document}